\documentclass[preprint]{aastex}
\usepackage{graphics,graphicx}
\usepackage[caption=false]{subfig}
\usepackage[]{natbib}
\usepackage[]{verbatim}
\usepackage[]{float}
\usepackage{gensymb}
\usepackage{latexsym,amsfonts,amsmath,amssymb}
\usepackage{lscape}
\newcommand{\upperRomanNumeral}[1]{\uppercase\expandafter{\romannumeral#1}}

\begin{document}
\title{The Effect of Binarity on Circumstellar Disk Evolution}

\author{Scott A. Barenfeld,\altaffilmark{1}
John M. Carpenter,\altaffilmark{2}
Anneila I. Sargent,\altaffilmark{1}
Aaron C. Rizzuto,\altaffilmark{3}
Adam L. Kraus,\altaffilmark{3}
Tiffany Meshkat,\altaffilmark{4,5}
Rachel L. Akeson,\altaffilmark{6}
Eric L. N. Jensen,\altaffilmark{7}
Sasha Hinkley\altaffilmark{8}
}
\altaffiltext{1}{California Institute of Technology, Department of Astronomy, MC 249-17, Pasadena, CA 91125, USA}
\altaffiltext{2}{Joint ALMA Observatory, Av. Alonso de C{\'o}rdova 3107, Vitacura, Santiago, Chile}
\altaffiltext{3}{The University of Texas at Austin, Department of Astronomy, Austin, TX 78712, USA}
\altaffiltext{4}{IPAC, Caltech, M/C 100-22, 1200 East California Boulevard, Pasadena, CA 91125, USA}
\altaffiltext{5}{Jet Propulsion Laboratory, California Institute of Technology, 4800 Oak Grove Drive, Pasadena, CA 91109, USA}
\altaffiltext{6}{IPAC-NExScI, Caltech, Pasadena, CA 91125, USA}
\altaffiltext{7}{Swarthmore College, Dept. of Physics \& Astronomy, 500 College Ave., Swarthmore, PA 19081-1390, USA}
\altaffiltext{8}{University of Exeter, Physics Department, Stocker Road, Exeter, EX4 4QL, UK}

\begin{abstract} 
We present new results on how the presence of stellar companions affects disk evolution based on a study of the 5-11 Myr old Upper Scorpius OB Association. 
Of the 50 G0-M3 Upper Sco members with disks in our sample, only seven host a stellar companion within 2$\arcsec$ and brighter than $K=15$, compared to 35 of 75 
members without disks. This matches a trend seen in the 1-2 Myr old Taurus region, where systems with a stellar companion within 40 au have a lower
fraction of infrared-identified disks than those without such companions, indicating shorter disk lifetimes 
in close multiple systems.  However, the fractions of disk systems with a stellar companion within 40 au 
match in Upper Sco and Taurus. 
Additionally, we see no difference in the millimeter brightnesses of disks in Upper Sco systems with and without companions, 
in contrast to Taurus where systems with a companion within 300 au are significantly fainter than wider and single systems.
These results suggest that the effects of stellar companions on disk lifetimes occur within the first 1-2 Myr of 
disk evolution, after which companions play little further role.   
By contrast, disks around single stars lose the millimeter-sized dust grains in their outer regions between ages of 1-2 Myr and 5-11 Myr. 
The end result of small dust disk sizes and faint millimeter luminosities is the same whether the disk has been truncated by a companion 
or has evolved through internal processes.

\end{abstract}

\section{Introduction} 
The formation and evolution of circumstellar disks is fundamental to our understanding of planet formation. 
This process 
begins with the collapse of a dense molecular cloud core and the subsequent formation of a protostar surrounded by an infalling envelope. Over a 
period of about 1 Myr, conservation of angular momentum causes the infalling material to form a circumstellar disk that remains around the star 
after the surrounding envelope is lost \citep[][and references therein]{Li2014}. This disk can provide the material for planet formation, a process that 
is not fully understood but likely involves direct collapse of disk material into a planet through gravitational instability and/or the slower growth of 
planetesimals and planets through core accretion \citep[e.g.,][]{Chabrier2014,Helled2014}. 
As the disk evolves, material will continue to viscously accrete onto the central star \citep[e.g.,][]{Hartmann1998}. 
At the same time, photoevaporation from the disk surface by high-energy stellar radiation dissipates disk 
material \citep{Owen2012,Alexander2014,Gorti2015}. Simultaneously, dust grains migrate inwards due to gas drag and grow to 
form larger bodies, depleting the small grain population \citep{Whipple1972,Weidenschilling1977,Brauer2007,Birnstiel2014,Testi2014}. By an age of 5-10 Myr, 
the majority of disks have dissipated \citep{Hernandez2008},
leaving behind a young star surrounded by any planets and associated debris that have formed.

Even for single stars, there are many uncertainties associated with the processes of disk evolution and planet formation. Additional complications arise 
from the fact that most stars are born in multiple systems. Studies of field stars show that the fraction 
of multiple systems is $\sim50\%$ among solar-type stars \citep{Raghavan2010} and $\sim30-40\%$ for later-type stars \citep{Fischer1992,Bergfors2010}. 
In the pre-main-sequence phase, multiplicity is at least as common \citep{Ratzka2005,Kraus2008,Lafreniere2008,Kraus2011,Cheetham2015}. Indeed, 
surveys of the earliest protostars indicate that a high binary fraction is intrinsic to the star formation process \citep{Chen2013}. 
Results from the \emph{Kepler} survey \citep{Borucki2010} show that while planet formation is suppressed in binary systems 
\citep{Wang2014a,Wang2014b,Wang2015a,Wang2015b,Kraus2016}, it is possible for such planets to form \citep[e.g.,][]{Holman1999,Dupuy2016,Hirsch2017}.
A complete understanding of the formation and evolution of stars and planets must therefore take the effects of stellar companions into account. 

Theoretical calculations have long predicted that the presence of a stellar companion will have an important influence on disk evolution \citep{Papaloizou1977}.
A disk around a single component of a binary system will be tidally truncated at approximately one-third to one-half of the binary separation and the resulting 
smaller disk will dissipate on a more rapid timescale than an unperturbed disk around a single star 
\citep[e.g.,][]{Artymowicz1994,Pichardo2005,JangCondell2015}. 
In fact, some initial surveys found that the fraction of binaries is lower in systems with 
disks and, in particular, accreting disks \citep{Ghez1993,Ratzka2005}, although other studies found no difference between accreting and non-accreting 
systems \citep{Leinert1993,Kohler1998}. 
Most recently, catalogs of much larger samples of disks identified with the \emph{Spitzer Space Telescope} \citep{Werner2004} and the 
\emph{Wide-field Infrared Survey Explorer} \citep[\emph{WISE;}][]{Wright2010} have provided more convincing evidence that the presence of stellar companions leads to shorter disk lifetimes \citep{Bouwman2006,Daemgen2016,Long2018}, 
with the disk fraction in 1-3 Myr old close binary systems ($\leq 40$ au separation) less than half that of wider binaries and single stars \citep{Cieza2009,Kraus2012,Cheetham2015}. 

Submillimeter interferometric observations are now providing high-resolution images of the outer regions of disks,
where most of the material resides, so that the effects of binarity on the entire disk, beyond the central regions probed by infrared observations, can be studied. 
In a millimeter study of 1-2 Myr old disks in Taurus, \citet{Harris2012} detected only one-third of disks in binary systems compared to two-thirds of 
single-star disks. In addition, the authors observed a positive correlation between binary separation and disk millimeter luminosity. 
While disks in binary systems with separations greater than 300 au had luminosities indistinguishable from single stars, 
disks in systems with a companion between 30 and 300 au were fainter by a factor of five. Disks in systems with a companion within 30 au were an 
additional factor of five fainter, implying that even in Taurus binary systems that maintain their disks, a substantial fraction of the millimeter-wavelength-emitting 
grains are lost due to the companion \citep[see also][]{Jensen1994,Jensen1996}.

Understanding how stellar companions affect later stages of disk evolution requires observations of older systems. Since these older disks are significantly fainter 
than their younger counterparts \citep{Nuernberger1997,Carpenter2002,Lee2011,Mathews2012,Williams2013,Carpenter2014,Ansdell2015,Barenfeld2016}, detailed studies require the 
sensitivity of the Atacama Large Millimeter/submillimeter Array (ALMA).  To this end, we measured the properties of over 100 
disks in the 5-11 Myr old Upper Scorpius OB Association (hereafter Upper Sco) using ALMA and found that these disks are a factor of $\sim4.5$ less massive \citep{Barenfeld2016} and a factor of $\sim3$ smaller \citep{Barenfeld2017} 
than their younger counterparts. In this paper, we consider how the influence of stellar companions has impacted the evolution of these disks to their current state. 
To investigate this, we searched for companions to the stars in our Upper Sco disk sample using adaptive optics (AO) imaging and aperture masking. We describe our sample,  
observations, and data reduction in Section \ref{sec:samp}. Section \ref{sec:id} specifies how companions were identified.  In Section \ref{sec:results}, 
we describe our detected companions and compare the companion frequency of systems with and without disks in Upper Sco. In Section \ref{sec:discussion}, 
we discuss how the effects of stellar multiplicity on disk properties vary with age in the context of disk evolution. Our conclusions are summarized in Section \ref{sec:summary}.

\section{Sample and Observations}
\label{sec:samp}
Our sample contains all 100 Upper Sco stars with spectral types between G2 and M4.75 (inclusive) as well as 13 M5 stars in Upper Sco
identified as hosting disks by \citet{Carpenter2006} and \citet{Luhman2012}.\footnote{Recent surveys, published after the present 
observations were obtained, have since expanded the known population of stars and disks in 
Upper Sco \citep{Esplin2018,Luhman2018}.} These disks were discovered based on excess 
infrared emission observed by \emph{Spitzer} and \emph{WISE} and include 82 disks classified as ``full,'' 
``evolved,'' or ``transitional'' by \citet{Luhman2012} based on their infrared colors. We consider these disks to be 
``primordial,'' i.e., a direct evolution of younger protoplanetary disks such as those in Taurus. The remaining 
31 disks in the sample are characterized as ``debris/evolved transitional'' \citep{Luhman2012}. These disks may represent the final 
phase of primordial disk evolution or be second-generation objects composed of dust created by the collision of 
planetesimals, with only an indirect evolutionary link to younger disks. 
The full sample is listed in Table \ref{tab:Disk_Sample} and receives a more detailed description in \citet{Barenfeld2016}. 
Distances to the stars in the sample are taken from the catalog of \citet{BailerJones2018}, inferred from \emph{Gaia} parallaxes using 
a Bayesian distance prior.

Twenty-seven systems in our sample have already been surveyed for stellar companions.  These systems are listed in Table \ref{tab:Lit_Comps}, along with the 
properties of any known companions. 
We obtained AO imaging and aperture masking observations 
of the remaining 86 stars using the NIRC2 AO imager (instrument PI: Keith Matthews) on the 10 m Keck II telescope. Targets were observed on the nights 
of 2011 May 15, 2013 May 30-31, and 2015 May 27-28.  Sources brighter than $R = 13.5$ were observed using natural guide star tip-tilt correction.  
Otherwise, a laser guide star was used \citep{Wizinowich2006}. 

Based on the \citet{Kraus2008} multiplicity survey of Upper Sco systems without disks identified by \citet{Luhman2012}, we expected to detect stellar companions 
at separations ranging from tens of milliarcseconds to several arcseconds. This range of separations can be probed using a combination 
of AO imaging, able to detect medium and wide separation companions, and nonredundant aperture masking, which achieves deeper contrast limits than AO imaging 
within a few hundred milliarcseconds. We thus observed our sample with both techniques using NIRC2.
Our observing procedure for each of these techniques is described below.

\subsection{Imaging Observations}
\label{sec:obs_imaging}
Our imaging observations are summarized in Table \ref{tab:Disk_Sample}.
For targets observed in 2013 and 2015, we acquired two 10 s AO images using either 
the $K'$ or $K_c$ filter on NIRC2. Targets with a Two Micron All Sky Survey 
\citep[2MASS;][]{Cutri2003,Skrutskie2006} magnitude brighter than $K_s=8.3$ were observed using the $K_c$ filter to prevent 
saturation. A third 10 s image was obtained of targets with a visually identifiable companion. If no such companion 
was seen, we obtained two further frames of 20 s AO images with the $K'$ filter.  These additional frames used 
a 600 mas diameter coronagraph for targets brighter than 2MASS $K_s = 10.6$ that would be partially visible behind the 
semi-transparent coronagraph.  Fainter targets were observed without the coronagraph, allowing us to easily determine 
primary positions when calculating companion separations. Due to unknown errors during observations, the two initial 10 s 
images were not saved on the nights of 2015 May 27-28, reducing the total integration times shown in Table \ref{tab:Disk_Sample}.
To avoid saturation in the initial and follow-up frames, 
we used shorter exposure times that were coadded to give the final 10 and 20 s frames. The exposure time 
per coadd was set based on the 2MASS $K_s$ magnitude of the target and the number of coadds was chosen to give total 
integration times of 10 or 20 s, respectively. 
Four targets, 2MASS J16070873-1927341, 2MASS J16071971-2020555, 2MASS J16073939-1917472, and 2MASS J16101473-1919095, 
were observed on 2011 May 15 as part of a separate program. For these targets, 10 frames of nine seconds each were obtained 
using the $K'$ filter without a coronagraph in place.

On the observing night of 2015 May 28, tip-tilt errors caused a number of targets to appear blurred in the images. For 
five of these sources, good quality observations from the previous night were available. For six sources, 2MASS J16020287-2236139, 2MASS J16050231-1941554, 
2MASS J16052459-1954419, 2MASS J16064102-2455489, 2MASS J16103956-1916524, and 2MASS J16124893-1800525, 
there is only data with poor tip-tilt correction. Despite these lower-quality data, we were still able to obtain useful detection limits for 
these systems in our comparison with other surveys (see Section \ref{sec:Sco_Comp}). For unknown reasons, 2MASS J16102819-1910444 was not visible in our images 
during observations. We exclude this source from our sample in the remainder of our analysis. 

The NIRC2 Preprocessing and Vortex Image Processing (VIP) packages\footnote{https://github.com/vortex-exoplanet/VIP} \citep{GomezGonzalez2017} were used to reduce the 
imaging observations. This included flat-fielding, dark subtraction, and bad pixel removal, as well as centering and de-rotation 
to align and stack individual frames for each target. High-order distortion corrections were applied using the 
solutions of \citet{Yelda2010} for the 2011 and 2013 data and the updated solutions of \citet{Service2016} for the 2015 data.

\subsection{Nonredundant Aperture Masking Observations}
\label{sec:obs_masking}
Nonredundant aperture masking observations 
were obtained if no obvious companion was revealed in the initial 10 s images.
We used a nine-hole mask with baselines ranging from 1.67 to 8.27 m. Images were read from a $512\times512$ pixel sub-array 
of the ALADDIN detector using multiple-correlated double sampling. We obtained six 20 s frames for each target observed in 2013, 
eight such frames for each target in 2015, and between 40 and 70 frames in 2011. Total integration times are given for each source in Table 
\ref{tab:Disk_Sample}. Depending on the brightness of the target, 
either 8, 16, or 64 endpoint reads were used along with coadds with shorter integration times in order to avoid saturation. 

Reduction of the aperture masking observations followed the procedure described in Kraus et al.\ 
(2008; see also Pravdo et al.\ 2006, Lloyd et al.\ 2006, Martinache et al.\ 2007, Kraus et al.\ 2011).\footnote{Reduction and analysis of the masking data were performed using the Sydney code (https://github.com/mikeireland/idlnrm).}
After dark-subtracting and flat-fielding, remaining bad pixels were removed from each frame. Frames were then spatially filtered using a 
super-Gaussian function of the form $\exp(-kx^4)$ to further reduce read noise.  Complex visibilities were extracted from Fourier transforms of 
the filtered frames.
To remove non-common path errors within the telescope and instrument, the data were calibrated using 
frames of Upper Sco targets that we determined were single.
Observations on the night of 2015 May 27 were taken with the telescope 
in position angle mode rather than vertical angle mode, causing the 
orientation of the nine-hole mask to change throughout the night and making this calibration more difficult. This led 
to shallower detection limits for these targets than those observed in vertical angle mode on other nights.

\section{Candidate Companion Identification}
\label{sec:id}

In this section we present how candidate companions were identified. We first describe the identification of astrophysical sources 
in our imaging and aperture masking data. We then discuss how the brightnesses and separations of these sources were 
used to determine whether or not they are likely to be physically associated companions. Finally, we identify 
potential wide-separation companions using \emph{Gaia}.

\subsection{Imaging}
\label{sec:Imaging_Analysis}

Stacked images of each of the 85 Upper Sco targets (excluding 2MASS J16102819-1910444) were searched for potential companions using VIP's \emph{detection} routine. 
These images were first convolved with the point spread function (PSF) of the primary star to enhance the signal of any potential companions. 
A two-dimensional Gaussian was then fit to local maxima of the unsmoothed 
image to compare the shape of the emission around each maximum to the expected PSF. For fits that displayed positive amplitude, 
had a center within two pixels of the location of the maximum, and had a full width at half maximum (FWHM) within three pixels of the PSF FWHM, the significance 
of the detection was determined by measuring its signal-to-noise ratio (S/N) in the unsmoothed 
image. The S/N was defined as
\begin{equation}
\label{eq:SNR}
S/N = \frac{F_{\mathrm{source}} - F_{\mathrm{bkg}}}{\sigma_{\mathrm{bkg}}\sqrt{1+\frac{1}{n}}},
\end{equation}
where $F_{\mathrm{source}}$ is the integrated flux of the source within one resolution element equal in diameter to the FWHM of the PSF. $F_{\mathrm{bkg}}$ and $\sigma_{\mathrm{bkg}}$ 
are the mean and standard deviation, respectively, of the integrated flux measured in resolution elements around an annulus at the radius 
of the potential source from the primary star. The number of these resolution elements within the annulus, $n$, corrects for the small-sample statistics introduced by the low number used \citep{Mawet2014}. Using this technique, we found 170 potential sources with an S/N greater 
than or equal to five. 

Subsequently, each image was inspected by eye to identify any speckles or other artifacts among detected sources that 
appeared at the 
same location in images of multiple targets. This inspection also 
located faint potential sources that the search algorithm missed due to, for example, another bright source or artifact 
at the same separation from the primary, which would increase the RMS noise at that separation. A total of 119 sources were 
rejected by this inspection, while 10 additional sources were identified.

Principal component analysis (PCA) using VIP was performed to subtract the stellar PSF and speckles from our images and improve 
our contrast limits \citep[e.g.,][]{Amara2012,Soummer2012}. Principal components were constructed from a PSF library composed of frames of other target stars 
found to be single by the above procedure. PCA was then applied to each target star, with the star itself excluded from the 
PSF library. We used 13 principal components and a library of 48 reference frames for images taken without the coronagraph. For images 
taken with the coronagraph, we used seven principal components and a library of 14 reference frames.
The above companion detection procedure was then repeated on the PSF-subtracted images.

In all, we identified 61 new sources from direct imaging that appear to be astrophysical but
may or may not be physically bound to the primaries. These detections are listed in Table \ref{tab:companions}. 
Relative photometry and astrometry of the sources in these systems were measured using the Python package \emph{photutils}
\citep{Bradley2016}.
The relative positions of primary stars and additional sources were derived from two-dimensional Gaussian 
fits. For targets with poor AO correction, centroids were estimated
using a ``center of mass'' technique that relied on the moments of a subimage around 
the source or primary.  Uncertainties on positions were estimated by measuring 
source locations in individual frames for each target and taking the maximum difference between any two 
frames.

Aperture photometry provided the relative fluxes 
of the primaries and additional sources with an aperture diameter equal to twice the FWHM of the 
primary.  
For systems with a detected source within 0\farcs3 of the primary, we used PSF-fitting photometry to measure the positions and relative fluxes.
PSFs were constructed with the algorithm described in \citet{Kraus2016}, which iteratively uses a library of single-star PSFs to generate 
template binary PSFs.

We estimated backgrounds and uncertainties 
in our aperture photometry using the mean and standard deviation of 20 apertures around an annulus at the same 
distance from the target source as the newly detected source. 
This accounts for both read noise and speckle noise, as well as any light from the primary that is included in our aperture photometry, 
as any such contamination will be incorporated into our background subtraction and uncertainties.
To measure background and uncertainties in the photometry of the primary stars, apertures were randomly positioned 
in annuli between $2\arcsec$ and $2\farcs5$ from the primary. For 2MASS J15562477-2225552, 2MASS J16020287-2236139, 2MASS J16020757-2257467, 
2MASS J16041740-1942287, 2MASS J16054540-2023088, 2MASS J16093558-1828232, and 2MASS J16220961-1953005, where sources lie within 
this separation range, annuli from $3\farcs5$ to $4\arcsec$ were used.

For our detected sources, photometric calibrations used the 2MASS \emph{K$_s$} magnitude of the primary and the ratio of integrated counts between each source
and primary. For systems with a source located within 
the $2\farcs6$ FWHM of the 2MASS PSF \citep{Skrutskie2006}, we separated out the \emph{K$_s$} magnitude attributable only 
to the primary. In addition to the photometric uncertainties described above, the uncertainties in new source 
magnitudes include the statistical uncertainty in the 2MASS magnitude of the primary and an assumed uncertainty 
of 0.05 magnitudes due to $K$-band variability of the primary \citep{Carpenter2001}.
Since the primary star is saturated in our images 
of 2MASS J16041740-1942287, 2MASS J16101888-2502325, and 2MASS J16154416-1921171, the \emph{K} magnitudes of the additional 
sources in these systems were determined using 
other targets observed during the same two-night runs to convert counts to \emph{K} 
magnitude. The separations and magnitudes of our newly detected sources are listed in Table \ref{tab:companions}.

Contrast limits are calculated for single stars using VIP's \emph{contrast curve} routine. This routine injects 
fake companions with a range of separations and contrasts relative to the primary into the stacked, PSF-subtracted frames for each target. 
The $5\sigma$ contrast limit is measured as the contrast of the brightest companion that is recovered with an S/N 
of less than five.  As above, noise is measured in the annulus at the angular separation of the fake companion using Equation \ref{eq:SNR}.
Our imaging contrast limits for sources without candidate companions (see Section \ref{sec:bound}) are listed in Table \ref{tab:limits}.

\subsection{Nonredundant Aperture Masking}
Nonredundant aperture masking achieves deeper contrast limits than traditional AO imaging at separations within a few hundred milliarcseconds 
using closure phases. At these separations, imaging contrast is limited by speckle noise created by 
atmospheric turbulence. This same turbulence introduces errors in the relative phases of the light reaching pairs of holes in the aperture mask.  
However, if these relative phases are summed around a triangle of the baselines 
connecting each pair, phase errors specific to individual holes, such as those due to atmospheric effects, will cancel out \citep[e.g.,][]{Lohmann1983,Readhead1988}. 
The resulting closure phases can then be used to search for close companions.

To locate companions in the aperture masking data, we adopted the technique used by \citet{Kraus2008}. 
Briefly, $\chi^2$ minimization was used to find the best-fit 
separation, contrast, and position angle of a potential companion for the closure phases of each target, along with the uncertainties 
in each of these parameters. The detection sensitivity to companions as a function of separation from the primary star was determined 
using 10,000 simulated data 
sets of a single star observed with the same $(u,v)$-sampling and closure phase errors as the observed data. 
The same fitting procedure was used to find the brightest detected companion in different annuli 
in each simulated data set. The detection threshold for each annulus was defined as the contrast ratio above which 
no potential companions were detected in 99.9\% of the simulated data sets. Table \ref{tab:companions} lists the six companions 
identified above this threshold. Table \ref{tab:limits} provides the contrast limits of the remaining targets.

\subsection{Selection of Candidate Companions}
\label{sec:bound}
\begin{figure}[!h]
\centerline{\includegraphics[scale=1.0]{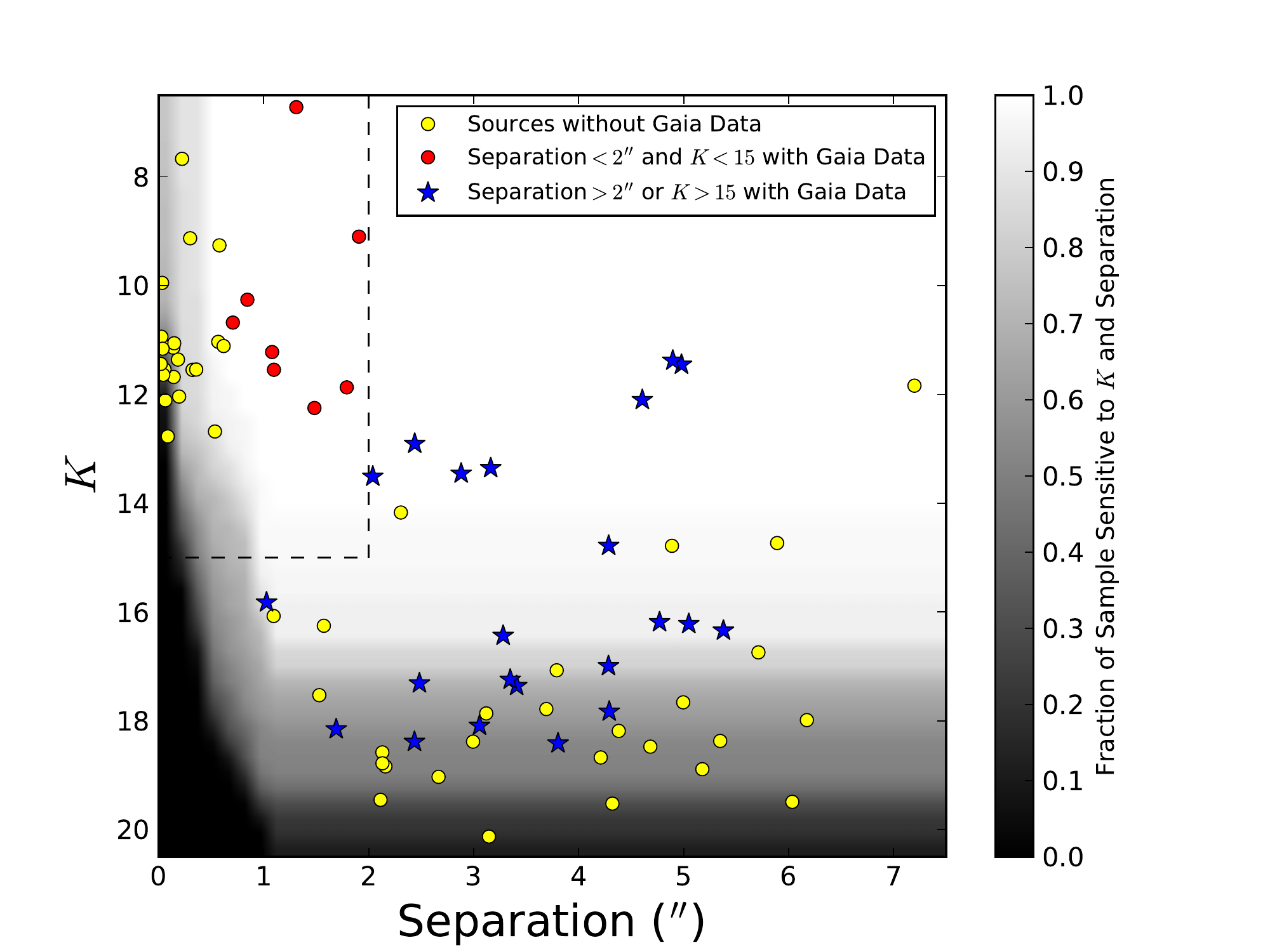}}
\caption{Projected separations and $K$ magnitudes of the 79 detected sources around Upper Sco 
stars with disks. We consider sources within $2\arcsec$ and brighter than $K=15$ to be candidate companions. 
This region is shown with dashed lines. 
A number of sources outside these limits may also be physically bound, but we expect significant background contamination among these 
sources. Red circles show sources that met our bound criteria and for which \emph{Gaia} data were available, while blue stars show sources 
with \emph{Gaia} data that did not meet our criteria. Sources with \emph{Gaia} data are also shown in Figure \ref{fig:Gaia}. The yellow circles show sources 
for which no \emph{Gaia} data were available. The grayscale background indicates the fraction of primary stars in the sample where the observations 
are sensitive to each 
$K$ magnitude and separation.
}
\label{fig:Kmags_Seps}
\end{figure}

The sources we detected are not necessarily bound companions to the host star. With only a single epoch of observations, we cannot use 
common proper motion to rule out the chance alignment of a field star. Instead, we use the 
brightness and separation of sources to distinguish between field stars and candidate companions.
Figure \ref{fig:Kmags_Seps} shows the $K$ magnitudes and separations of the 67 sources found by 
imaging and masking and the 12 literature companions listed in Table \ref{tab:Lit_Comps}. 
We used the TRILEGAL galactic population models \citep{Girardi2005} to simulate the population of 
field stars as a function of $K$ magnitude in the direction of Upper Sco. We find a density of 
$2.2\times10^{-4}$ field objects per square arcsecond brighter than $K=15$. For 
our full sample of 112 targets, 
we would expect a total of less than one such field star to be 
within 2$\arcsec$ of a target star by chance. We therefore consider any sources brighter than $K=15$ and 
within 2$\arcsec$ of a target star likely to be a candidate bound companion. These limits are the same as those used in 
\citet{Kraus2008} to identify candidate companions in Upper Sco and are shown in Figure \ref{fig:Kmags_Seps} 
as dashed lines. Sources that meet these criteria are indicated in the ``Candidate Companion'' column of Table 
\ref{tab:companions}.
For consistency, we apply these criteria to the previously known 
companions in Table \ref{tab:Lit_Comps}, even if objects beyond these limits have been confirmed to be associated by other methods.

\begin{figure}[!h]
\centerline{\includegraphics[scale=1.0]{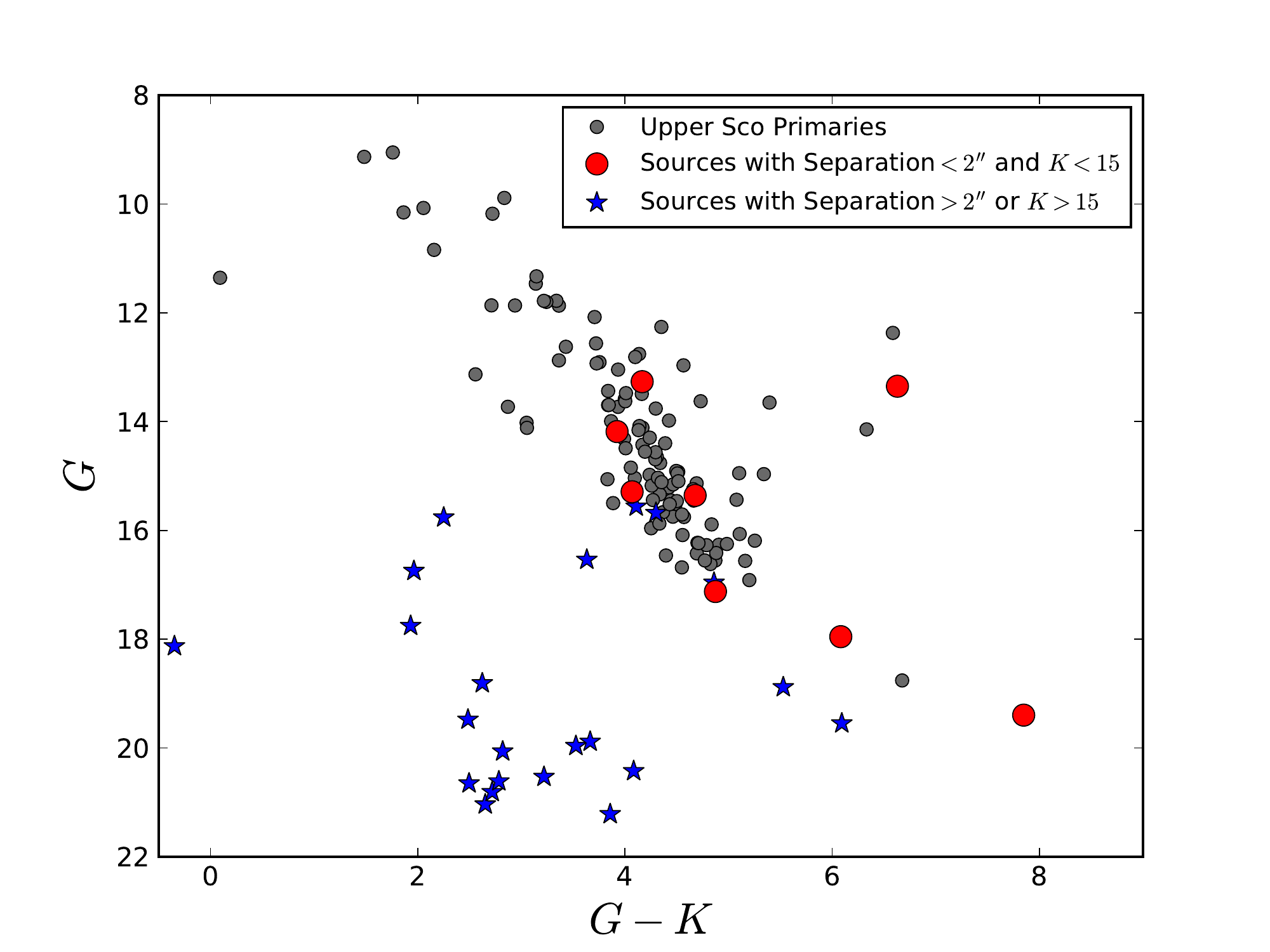}}
\caption{Color-magnitude diagram of Upper Sco primaries in our sample (gray points) and additional sources that 
meet (red circles) and fail to meet (blue stars) our criteria of separation $<2\arcsec$ and $K<15$ to be considered 
candidate companions. Sources that meet our criteria lie along the same color-magnitude sequence as the Upper Sco primaries, as expected. 
Sources outside of these criteria are typically bluer and fainter than this sequence, consistent 
with background stars.
}
\label{fig:Gaia}
\end{figure}

Figure \ref{fig:Gaia} presents the color-magnitude diagram for sources in the \emph{Gaia} DR2 Catalog \citep{GAIA,GAIADR2}. 
The candidate companions that meet our criteria for physical association lie along the same sequence as 
the primary stars, as would be expected for co-evolutionary companions at the same distance from Earth. 
The sources that do not meet these criteria include a small number of objects that match the 
colors and magnitudes of the candidate companions and primaries.  However,
the majority of objects outside of our selection criteria are fainter and bluer than the primary star sequence, 
as would be expected for background field stars. While we cannot rule out that a fraction of sources 
fainter than $K=15$ and separated by more than 2$\arcsec$ are physically associated companions, there is a significant 
fraction of field objects beyond these limits. 

We note that the sources beyond $2\arcsec$ that are fainter than $K\sim 12.5-13$ would be candidate brown dwarfs ($M\lesssim0.08$ M$_{\odot}$) if they were associated, assuming a distance and 
age of 145 pc and 5-10 Myr \citep{Chabrier2000,Baraffe2002}. Similarly, sources fainter than $K\sim15.5-16$ would be 
potential giant planets ($M\lesssim13$ M$_{\mathrm{Jup}}$) if they were bound. While these objects are most likely field stars, they may be worth observing in the future to look 
for common proper motion.

\subsection{Candidate Wide Companions with \emph{Gaia}}
\label{sec:Gaia}
\begin{figure}[!h]
\centering
\subfloat{\includegraphics[width=0.50\linewidth]{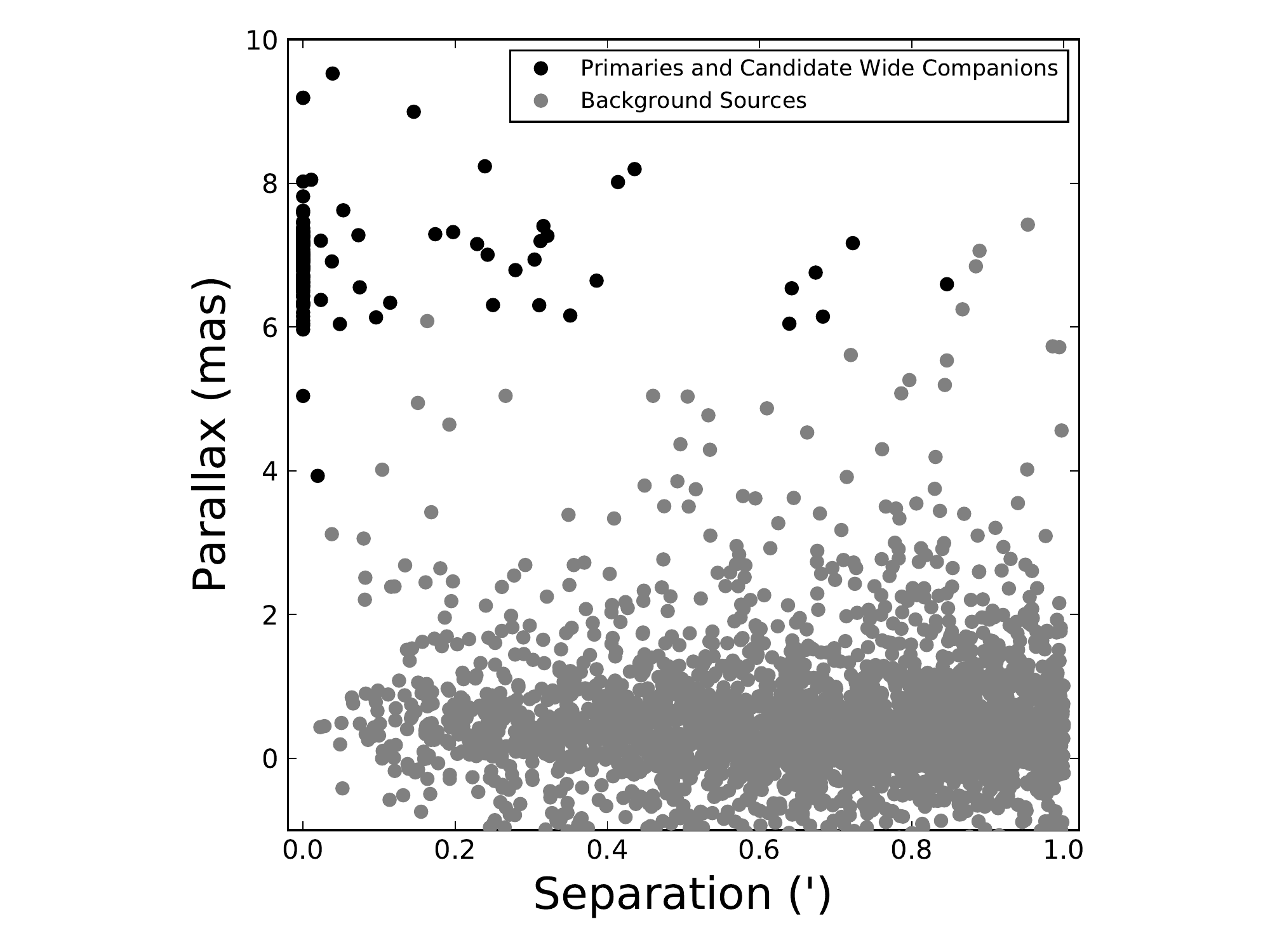}}
\subfloat{\includegraphics[width=0.50\linewidth]{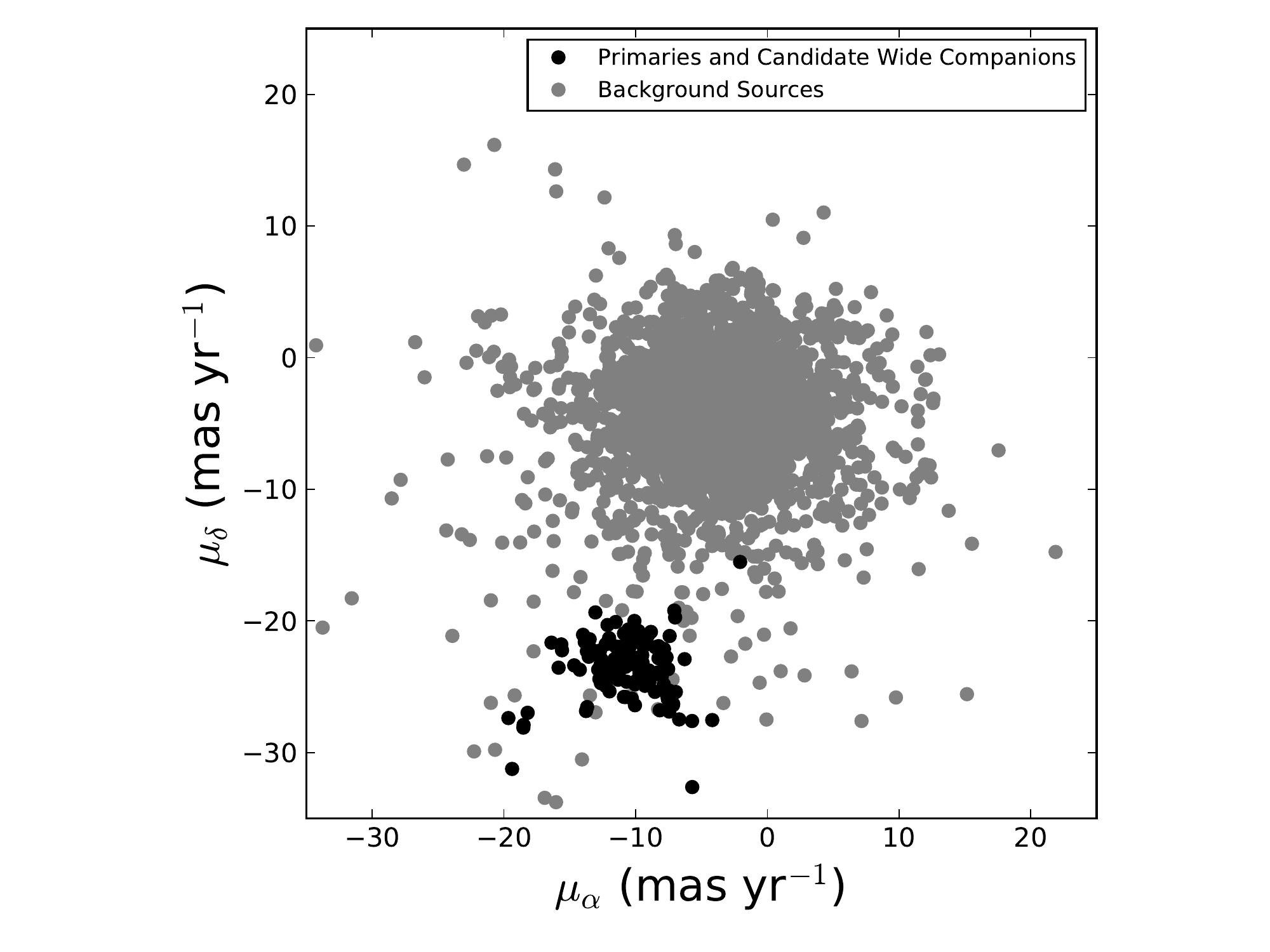}}
\caption{Parallaxes (left) and proper motions (right) of all sources in the \emph{Gaia} DR2 Catalog 
within $1\arcmin$ of the targets in our Upper Sco disk sample. Most sources have parallaxes and proper 
motions close to zero, as expected for background objects. The black points show the primaries and 
candidate wide companions. The candidate wide companions have parallaxes and proper motions 
similar to their primaries and clearly distinct from the background sources.
}
\label{fig:wide_companions}
\end{figure}

To search for potential companions at wider projected separations, we used the \emph{Gaia} DR2 Catalog to
identify any sources within 1$\arcmin$ of a target star in our Upper Sco disk sample. Figure \ref{fig:wide_companions} shows the 
\emph{Gaia} parallaxes and proper motions of these sources. The majority of sources have parallaxes and proper motions concentrated 
close to zero, as expected for background objects. For each primary star in the sample, we searched for any additional 
sources with similar parallax and proper motions that stood out from the background sources. Figure \ref{fig:wide_companions} 
shows these candidate wide companions and primaries, which are clearly separated from the main cluster of background objects.
These sources, listed in Table \ref{tab:Gaia_Companions}, have parallaxes within three milliarcseconds of 
their potential primaries and proper motions in R.A. and decl. within five milliarcseconds per year.

\section{Disks and Multiplicity in Upper Sco}
\label{sec:results}
In this section, we describe the Upper Sco candidate companions discovered in our survey. We determine the 
locations of the millimeter disks in these systems relative to the primary and companion(s). 
We then compare the companion fractions of stars with and without disks in Upper Sco.

\subsection{Properties of Upper Sco Systems with Disks and Companions}
\begin{figure}[!h]
\centerline{\includegraphics[scale=1.0]{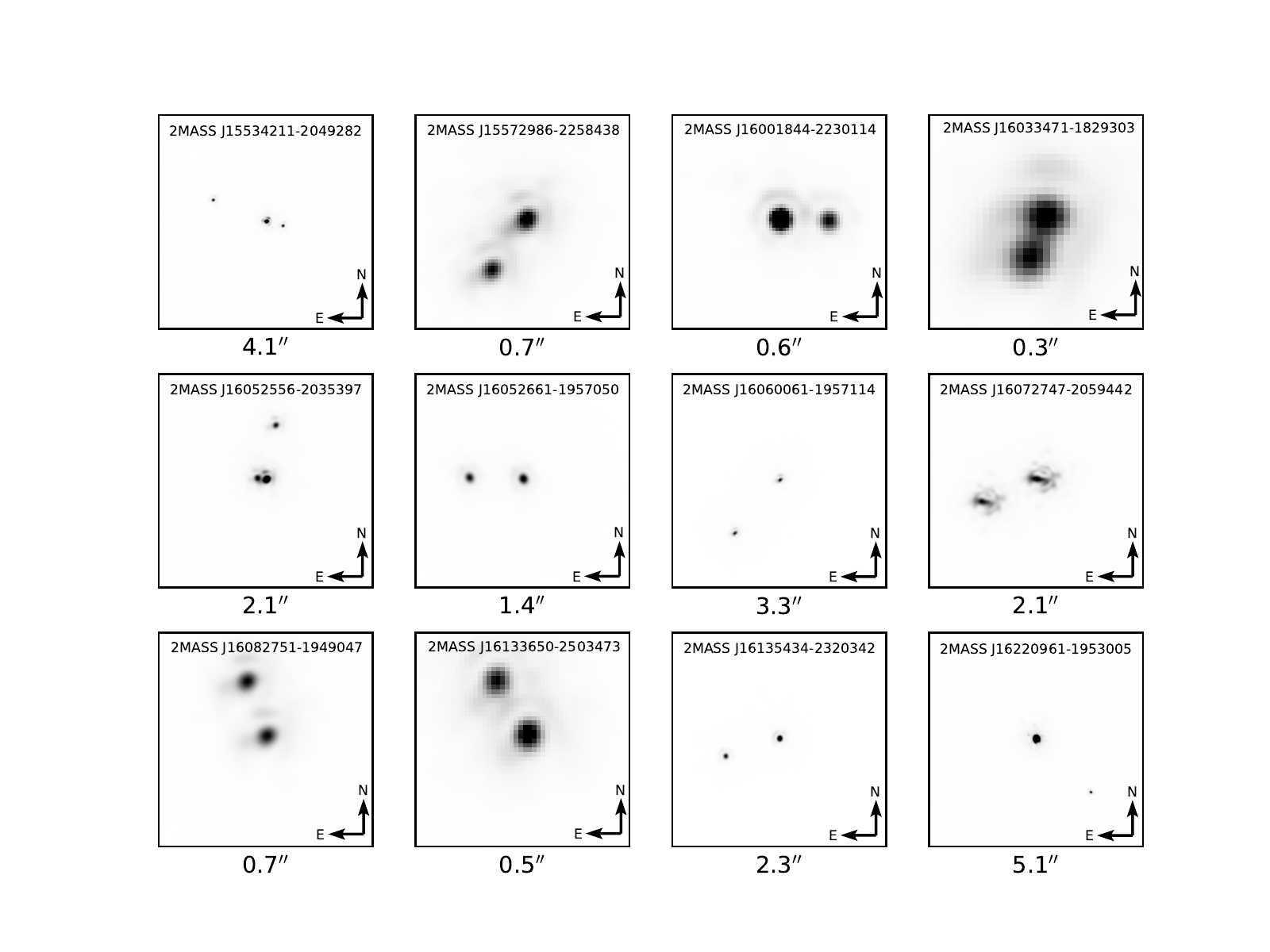}}
\caption{NIRC2 $K'$ images of the Upper Sco disk systems with new companions 
discovered by imaging in this survey. The angular extent of each image is indicated 
for each panel.
}
\label{fig:NIRC2_Images}
\end{figure}

We found 30 candidate companions in 27 systems brighter than $K=15$ and with separations of less than 2$\arcsec$. 
These includes the previously known companions listed in Table \ref{tab:Lit_Comps} that meet these criteria. 
Newly discovered candidates are indicated 
in Table \ref{tab:companions} by the ``Candidate Companion'' column.
Of the 81 primordial disk systems in the sample, 22 contain a candidate companion, along with five of the 31 debris/evolved transitional disks. 
The companions range in
separation from 0\farcs02 to 1\farcs91, corresponding to projected separations of 2.8-265 au assuming the distances 
listed in Table \ref{tab:Disk_Sample}. $K$-band 
magnitudes of these objects range from 6.72 to 12.77. NIRC2 $K'$  
images of the 12 systems with new companions discovered by imaging are shown in Figure \ref{fig:NIRC2_Images}. Ten 
of these systems include a single candidate companion, while two targets, 2MASS J15534211-2049282 and 2MASS J16052556-2035397, 
appear to be triple systems.

Figure \ref{fig:ALMA_Images} shows ALMA 880 $\mu$m continuum images of the 26 systems with companions for which we 
have ALMA data \citep{Barenfeld2016}. These exclude 2MASS J16033471-1829303, an M5 star with a disk identified by infrared excess \citep{Luhman2012} 
that was not observed with ALMA. The relative positions of the primary and companion(s) are overlaid in each image. 
The locations of the primary stars at the time of the ALMA observations were calculated 
using positions and proper motions from the \emph{Gaia} DR2 Catalog. When \emph{Gaia} proper motions or positions were unavailable, 
we used data from the PPMXL catalog \citep{Roeser2010}.
For 16 systems, 
the millimeter emission is only at the location of the primary star or is not detected toward either component.  Individual disks 
are detected around each component of 2MASS J16113134-1838259 and 2MASS J16135434-2320342. The disk 
in 2MASS J16052556-2035397 appears to be located around the wider companion of this triple system. 
This may also be the case for 2MASS J16082751-1949047. However, the uncertainties of the R.A. and decl. of the primary 
star are $0\farcs11$ due to only data from PPMXL being available 
for this system. We therefore cannot definitively determine the relative positions of the disk and stars. 
Six other systems, 2MASS J15534211-2049282, 2MASS J16001844-2230114, 
2MASS J16043916-1942459, 2MASS J16075796-2040087, 2MASS J16133650-2503473, and 
2MASS J16141107-2305362 show disk millimeter emission that encompasses both stellar components at the resolution of the ALMA observations. 
The disks in 2MASS J16082751-1949047 and these six other systems may exist around one or both stars individually or may be circumbinary.

Figure \ref{fig:SEDs} shows the infrared spectral energy distributions (SEDs) of the seven systems where the millimeter-wavelength emission cannot be 
conclusively assigned to the primary or secondary given the angular resolution of the ALMA observations. 
Infrared photometry is from 2MASS \citep{Cutri2003}, \emph{Spitzer}, and \emph{WISE} \citep{Luhman2012}. Stellar photospheres were estimated assuming blackbody emission 
with the same stellar parameters as in \citet{Barenfeld2016}. Six systems show infrared excess 
at wavelengths shorter than 10 $\mu$m, indicating the presence of warm dust. This does not necessarily rule out circumbinary 
disks, but we can say that there must be dust around one or both individual stars. Since 2MASS J16043916-1942459 
exhibits an infrared excess only at 24 $\mu$m and has a companion with a projected separation of only 3.8 au, this system 
is likely to be a circumbinary disk. However, given the weakness of the 24 $\mu$m excess and low S/N 
of the ALMA image, its nature is difficult to determine with certainty.

\subsection{A Comparison of Upper Sco Systems with and without Disks}
\label{sec:Sco_Comp}
We now compare the stellar companion fraction for Upper Sco stars with and without circumstellar disks.
As described in Section \ref{sec:id}, we have detected 30 candidate companions brighter than $K=15$ and with separations 
of less than $2\arcsec$ in 27 of 112 systems with 
disks identified from infrared colors (see Section \ref{sec:samp}. Our comparison sample is composed of the 77 Upper Sco stars without such disks surveyed for 
stellar companions by \citet{Kraus2008} using similar observations to those presented here. This sample, listed in Table \ref{tab:Kraus_Sample}, ranges in 
spectral type from G0 to M4 (inclusive) and is described in detail by \citet{Kraus2008}. Companions 
identified in this sample meet the same brightness and separation criteria used in this work.

\begin{figure}[!h]
\centerline{\includegraphics[scale=1.0]{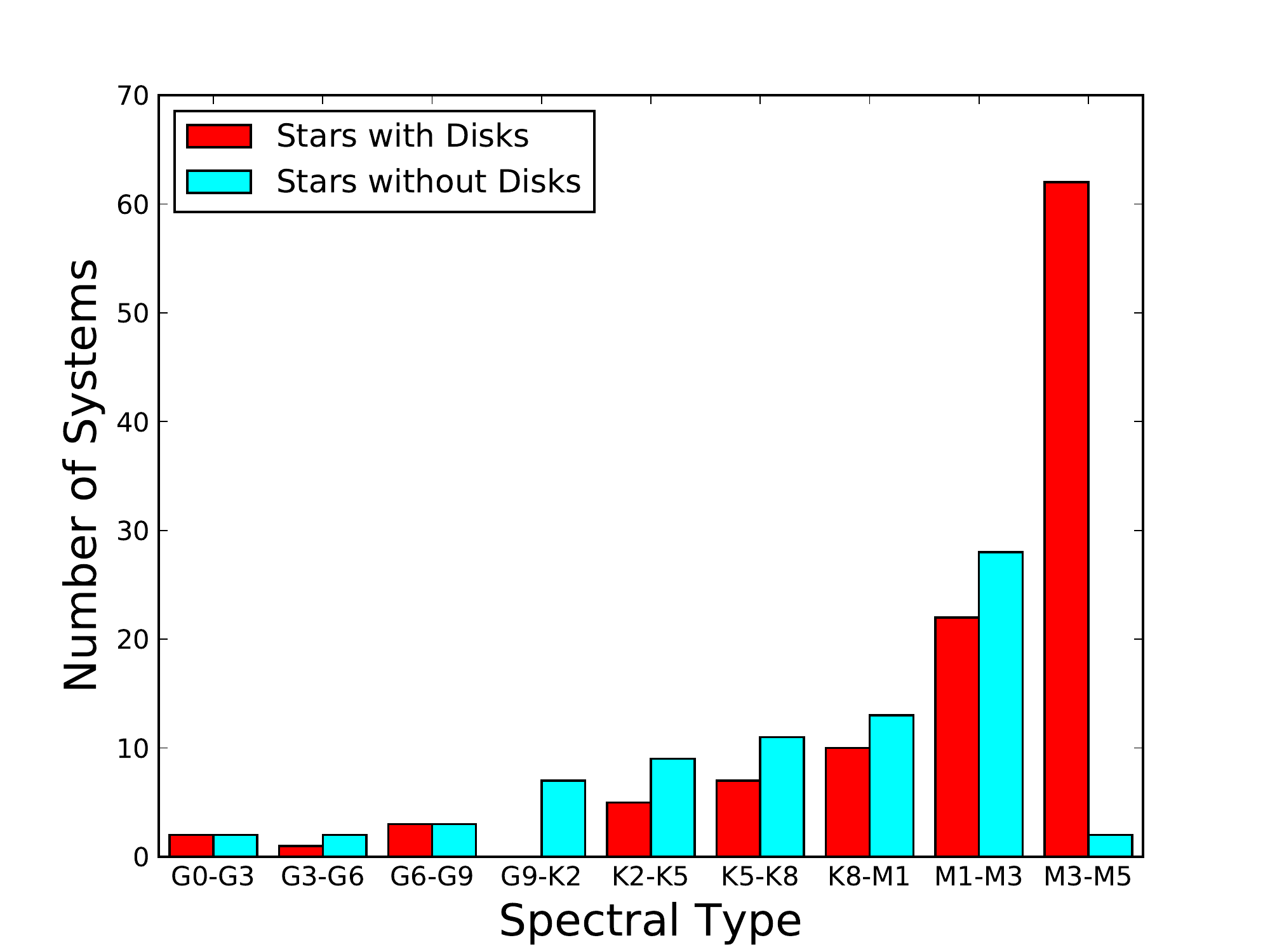}}
\caption{Spectral-type distributions of Upper Sco primary stars with (red) and without (cyan) 
disks. This disk sample includes 62 systems with spectral types later than M3, compared to only 
two such systems without disks. Restricting to spectral types M3 and earlier, the
samples are consistent with being drawn from the same parent distribution.
}
\label{fig:SpTs}
\end{figure}

To ensure a meaningful comparison of systems with and without disks, we examined the spectral-type distributions 
of these samples.  The distributions of primary star spectral types for the two samples are shown in Figure \ref{fig:SpTs}. 
Only two of the 77 systems without disks have spectral types later than M3, compared to 62 of the 112 systems in 
the disk sample. The latter sample was extended to later spectral types in order to include a larger number 
of Upper Sco systems with disks in the studies by \citet{Barenfeld2016} and \citet{Barenfeld2017}. 
Given the lack of M4 and M5 stars in the \citet{Kraus2008} sample, we restrict our comparison of companion fractions 
to systems with primary spectral types of M3 or earlier.
With this restriction, the spectral types of the two samples are consistent with being drawn from the same distribution, 
with a $p$-value of 0.17, 
according to the $\chi^{2}$ test implemented with the \emph{R Project for Statistical Computing} \citep{R}. 
This result is independent of how the spectral types are categorically binned.

We note that, while similar techniques were used to observe the disk and comparison samples, different observing conditions 
may have led to discrepancies in the sensitivity to companions between the two samples.  In addition, literature data that did not 
include aperture masking was used for several systems in the disk sample, reducing our sensitivity to close-in companions relative to 
the comparison sample. We estimate below the number of companions this may have caused us to miss in the disk sample.

\begin{figure}[!h]
\centerline{\includegraphics[scale=1.0]{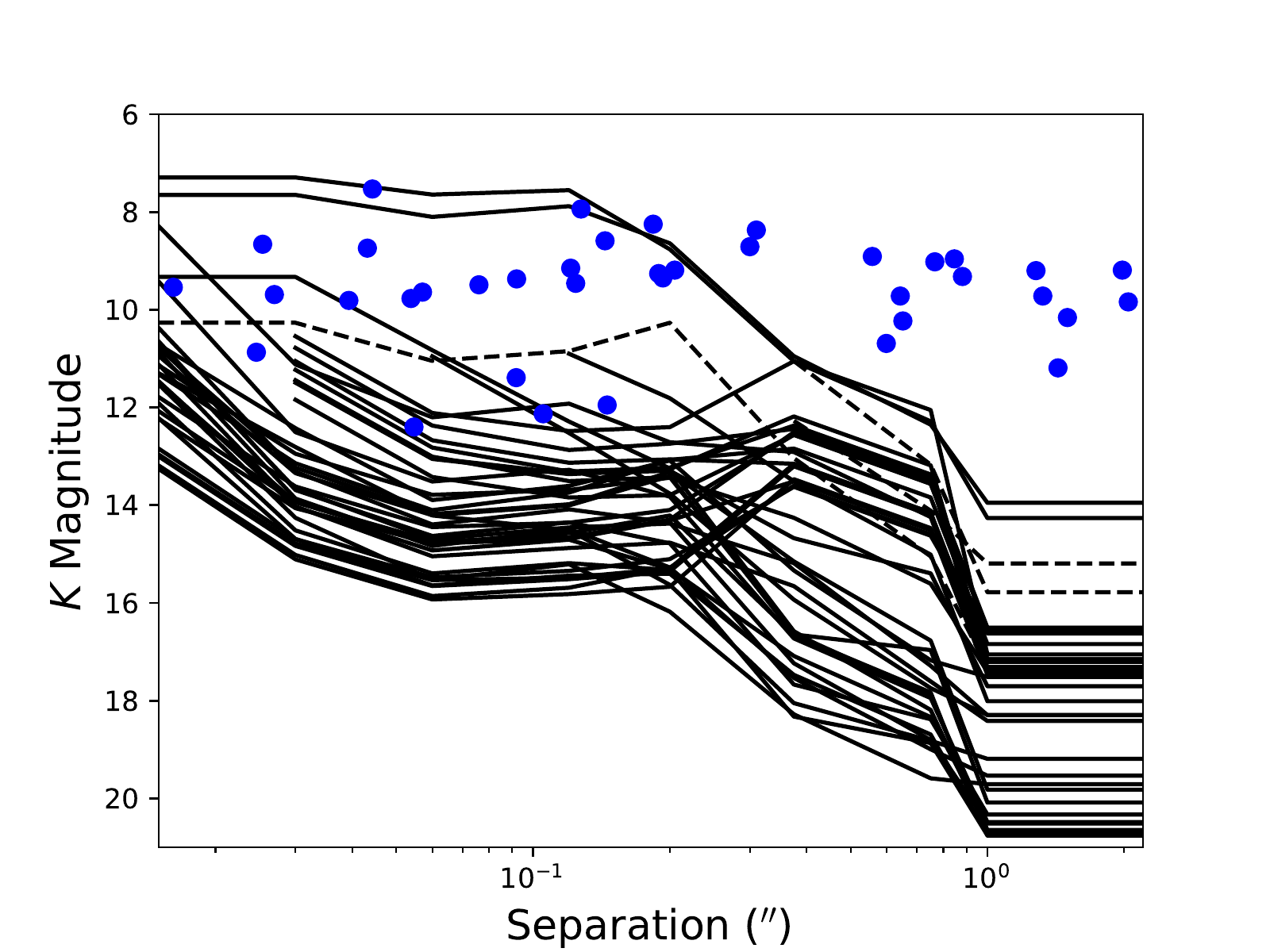}}
\caption{Apparent magnitude detection limits as a function of separation for 
Upper Sco disk-hosts with no candidate companions and spectral types of M3 or earlier. 
The dashed curves show the contrast limits for the three sources with 
poor tip-tilt in this spectral type range. 
The blue points show 
the companions found by \citet{Kraus2008} among a sample of Upper Sco 
stars without disks. The majority of observations in the current disk 
sample were sensitive enough to have detected all of these companions 
if they were present around the disk-hosting stars. Under the assumption 
that the stars with disks have the same population of companions as those 
without disks, we would have only expected to miss approximately two to three of these companions 
due to lower sensitivities. 
}
\label{fig:limits}
\end{figure}

Our aim in this study was to determine whether the fraction of disk systems with a stellar companion is lower than that of systems without disks. Thus, 
to compare survey completeness, we estimated the number of companions detected in diskless systems that would have been missed 
if they existed with the same brightness and separation around stars in the disk sample.
Figure \ref{fig:limits} shows the limiting magnitude as a function of separation of the disk systems for which 
no companion was found. Also plotted are the magnitudes and separations of the companions found in the diskless 
sample for systems with primary spectral type M3 or earlier. The majority of these companions would have been detected had they existed around the stars in our disk 
sample. The companions that may have been missed were found using aperture masking by \citet{Kraus2008}. 
Our sensitivities to these close-in sources are lower for a number of stars in our sample due to masking data 
not being available, calibration issues due to data being taken in position angle mode, and tip-tilt correction 
problems (see Section \ref{sec:samp}). For example, 2MASS J16142029-1906481 was observed without masking by \citet{Lafreniere2014}. 
If the 75 systems in the diskless sample had been observed with the same sensitivity achieved for this source, 
companions detected by \citet{Kraus2008} would have been missed in 14 systems, equal to $19\%$ of the diskless sample.
If 2MASS J16142029-1906481 followed the same underlying companion probability distribution as 
the diskless sample, we would thus have expected to miss 0.19 companions on average.
Similarly, our observations of 2MASS J16103956-1916524, which suffered from poor tip-tilt correction, 
would not have detected five companions from the diskless sample for an expected value of 0.07 companions missed. 
The fraction of \citet{Kraus2008} companions in systems without disks 
that would have been missed in our disk sample 
can be calculated in this manner for each star in the sample. With this calculation, we found that even if systems with and without 
disks shared the same distribution of companion brightnesses and separations, we would have only expected to not detect approximately two to three 
companions in the disk sample due to lower sensitivities.
Restricting ourselves to the primordial disks in our sample, we would have expected to miss fewer than one companion relative to the diskless sample.

With this caveat in mind, we now compare the companion fractions of Upper Sco systems with and without disks. 
For spectral types M3 and earlier, 35 out of 75 stars without disks 
have at least one companion. By contrast, only seven out of 50 systems with disks include companions. From the Fisher Exact Test, 
the probability that the lower companion fraction in star-disk systems is due to chance is $2\times10^{-4}$.
Even if our previous estimate of three missed companions were added to the total number of companions observed around stars with disks, 
the Fisher Exact Test would still give a probability of $2\times10^{-3}$ that the companion fractions are the same for stars with and without 
disks.
Since this includes the debris/evolved transitional disks and we are primarily concerned with the 
evolution of primordial disks, we eliminated the potential debris disks and repeated the comparison. We found that
six of the 26 
primordial disk systems with spectral types M3 and earlier host companions, giving a $p$-value of 0.04 when compared to 
the stars without disks. 
Thus, the fraction of multiple systems among stars with primordial disks is lower than that of stars without disks 
with marginal significance.

\citet{Kuruwita2018} have also studied the effect of binarity on the presence of disks in Upper Sco in a radial velocity search 
for stellar companions to 55 Upper Sco G, K, and M stars with an infrared excess. The authors find a stellar companion fraction for these systems 
of $0.06^{+0.07}_{-0.02}$ for periods less than 20 years. This is lower than the fraction expected for field stars with the same primary mass distribution, 
$0.12^{+0.02}_{-0.01}$, although the fractions agree within uncertainties. This survey probes separations within $\sim0\farcs05$ at the $\sim145$ pc distance of Upper 
Sco, separations similar to and within the inner working angle of our current aperture masking observations. Thus, it would be possible with a larger radial-velocity sample 
to test if the lower companion fraction in systems with disks relative to those without disks found in the present study holds for closer-separation companions. 
Such a sample was recently provided by \citet{Esplin2018}, who compiled an updated census of 484 Upper Sco disks identified by infrared excess.

\subsection{2MASS J16075796-2040087: An Accreting Circumbinary Disk}
\label{sec:160757}
While the majority of the disks in the Upper Sco multiple systems in our sample appear to be located around a single star within each system, 
the disk in 2MASS J16075796-2040087 is likely to be circumbinary. This 
system has a stellar companion at a projected separation of 6.3 au and a disk with 880 $\mu$m flux density of 23.49 mJy, 
one of the brighter millimeter sources in the present sample. Corrected for the updated \emph{Gaia} distance to this system in Table \ref{tab:Disk_Sample}, 
\citet{Barenfeld2017} found that the dust disk in this system 
extends to $15\pm1$ au while the gas component reaches to $46^{+6}_{-2}$ au, well beyond the projected companion separation. 
While it is possible that the physical separation of the components of this system is wider than their projected separation, 
it would have to be over a factor of seven larger to be outside of the gas disk. \citet{Harris2012} constructed 
the probability distribution for the ratio of physical to projected separation of a binary using a Monte Carlo simulation of 
the underlying orbital parameters. Depending on the assumed priors for orbital parameters, the distribution peaks between a ratio 
of 0.5 and 1.5, with only a low probability tail extending beyond a ratio of 3. 2MASS J16075796-2040087 is therefore most likely to be a 
circumbinary disk.

However, in
Figure \ref{fig:SEDs}, there is a strong infrared excess at wavelengths as short as 1.7 $\mu$m, 
indicating the presence of hot dust close to one or both of the stars. We note that the stellar photospheric emission
calculated for this system assumes a spectral type of M1 \citep{Luhman2012}, while the primary star may have an earlier spectral type 
\citep[see][]{Kraus2009,Cody2017}. Despite the uncertainty in the stellar photosphere, it is clear that there 
is significant circumstellar material around at least one of the stars in this system. \citet{Kraus2009} found that 
there is likely to be an accretion-powered outflow based on strong optical emission lines, while 
\citet{Cody2017} observed bursting behavior on a $\sim15$ day timescale in the optical light curve, consistent with 
episodic accretion.

One possible explanation for these observations is that material from the inner edge of the circumbinary disk is streaming 
across the dynamically cleared inner gap and accreting onto one or both of the stars \citep[e.g.,][]{Artymowicz1996,Gunther2002}.
The details of this process depend strongly on the mass ratio and orbital parameters of the binary, but it is 
generally expected that this accretion will be modulated with a period of order that of the binary orbit \citep{Munoz2016}. 
Modulated accretion has been observed in spectroscopic binaries with circumbinary disks such as DQ Tau \citep{Mathieu1997}, UZ Tau E \citep{Jensen2007}, 
and TWA 3A \citep{Tofflemire2017a,Tofflemire2017b}.
However, 2MASS J16075796-2040087 exhibits optical variability on a $\sim15$ day timescale, much shorter 
than the orbital period of a binary with a projected separation of 4.6 au. Direct accretion onto 
the stars in the binary is only expected for spectroscopic binaries with separations of a fraction of an au. 
In wider systems, inner circumprimary and circumsecondary disks are expected to be fed and maintained by the streams \citep{Gunther2002,Dutrey2016}. 
Observations of GG Tau \citep{Dutrey1994,Dutrey2014}, with a projected separation of $\sim35$ au, UY Aur \citep{Close1998,Duvert1998,Tang2014},  
$\sim125$ au, and L1551 \citep{Takakuwa2014}, $\sim70$ au, fit such a scenario. A similar process 
may be taking place in 2MASS J16075796-2040087. Though the 4.6 au binary separation makes this system
an intermediate case between spectroscopic binaries and wider pairs such as GG Tau, a circumprimary 
and/or circumsecondary disk replenished by streams from the outer circumbinary disk may be present. 
Accretion from the inner disk(s) may then be causing the observed optical emission lines, infrared excess, and variability on 
timescales unrelated to the binary orbital period.

\section{Discussion}
\label{sec:discussion}

In this section we investigate how the relationship between disks and stellar companions varies with age.
We compare the fractions of disk systems with close companions and examine the relationship between companion separation 
and disk millimeter luminosity in the 1-2 Myr old Taurus and 5-11 Myr old Upper Sco regions. We then discuss the implications 
of these results for disk evolution.

\subsection{Companion Frequency of Disk Systems in Taurus and Upper Sco}
\label{sec:Taurus_Comp}
Studies of how disks are affected by stellar companions in Taurus and other young star forming regions have shown that multiplicity has a significant 
impact during the first 1-2 Myr of disk evolution. The infrared-detected disk fraction of 1-2 Myr old stars with close companions ($\leq 40$ au separation) is 
lower by approximately a factor of two to three than that of single stars of the same age \citep{Cieza2009,Kraus2012,Cheetham2015}. 
In Upper Sco (age 5-11 Myr), infrared-detected disks are also less frequent for systems with a close companion than 
for single stars, but by approximately the same factor of two to three seen for 1-2 Myr old systems \citep{Kraus2012}. This suggests that after the first 1-2 Myr 
of a disk's evolution, the presence of a companion has no further effect on disk frequency as traced by dust infrared emission.

We tested the effect of stellar companions on disks between the ages of Taurus and Upper Sco using the expanded sample of Upper Sco binaries presented in this work. Our sample was specifically 
chosen to include Upper Sco systems with infrared-detected disks. Due to this selection criterion, we could not compare the disk frequencies of close binaries to that 
of single stars.  Instead, we compared the fraction of close companions among systems with disks in Taurus and Upper Sco. 
Of the 83 Taurus G, K, and M stars with infrared-detected disks listed in \citet{Kraus2012} that have been surveyed for companions, 13 host 
a stellar companion within a projected separation of 40 au. In the present Upper Sco survey, we find 11 stars with such companions among the 82 
primordial infrared-detected disks in our sample. These close companion fractions are consistent according to the Fisher exact test, with a 
$p$-value of 0.83.  This supports the \citet{Kraus2012} result that stellar companions have little to 
no effect on disk evolution as traced by infrared-emitting dust after the first 1-2 Myr. Instead, the lower companion fraction for systems with infrared-detected disks 
in Upper Sco relative to those without disks (Section \ref{sec:Sco_Comp}) is simply due to the reduction in the disk fraction 
of multiple systems that occurs before an age of 1-2 Myr.

\subsection{Millimeter Emission and Multiplicity} 
\label{sec:Harris}

\citet{Harris2012} found a clear relationship between companion separation and disk millimeter luminosity in Taurus multiple 
systems. Taurus disks in systems with projected companion 
separations between 30 and 300 au are fainter by a factor of five than those in single-star and wider-companion systems, while disks in systems with companions projected within 30 au are an 
additional factor of five fainter. We now use the current sample to test this relationship in Upper Sco and compare the results to Taurus.

Our goal was to isolate the effect of binarity on disk evolution. For Upper Sco, we used the 
Upper Sco primordial disk systems in the current sample with ALMA 0.88 mm continuum flux density 
measurements from \citet{Barenfeld2016}. For Taurus, we used the compilation of 1.3 millimeter flux densities of infrared-identified 
Class II Taurus systems from \citet{Akeson2019} and selected systems classified as primordial disks by \citet{Luhman2010}. 
Flux densities, originally measured by \citet{Andrews2013}, \citet{Akeson2014}, \citet{WardDuong2018}, and \citet{Akeson2019}, 
have been scaled to 0.88 mm using the scaling factor of 2.55 assumed for Taurus 
disks by \citet{Andrews2013}. We restricted the Taurus sample to systems with single or primary stellar mass between 0.14 M$_{\odot}$ and 1.7 M$_{\odot}$ 
to match the stellar mass range of the Upper Sco sample \citep{Barenfeld2016}. Within this range, 78\% of Taurus systems in our final comparison 
sample have a single star or primary stellar 
mass below 0.6 M$_{\odot}$, compared to 87\% of Upper Sco systems. We note, however, that the Upper Sco sample is skewed toward slightly lower stellar masses 
than that of Taurus, with 69\% of systems $<0.3$ M$_{\odot}$ compared to 24\% in Taurus. For both samples, we excluded triple and higher-order systems in order to 
isolate the effect of a single companion separation. We also excluded circumbinary disks to focus on the effects of disk truncation 
by an external companion.

\begin{figure}[!h]
\centerline{\includegraphics[scale=1.0]{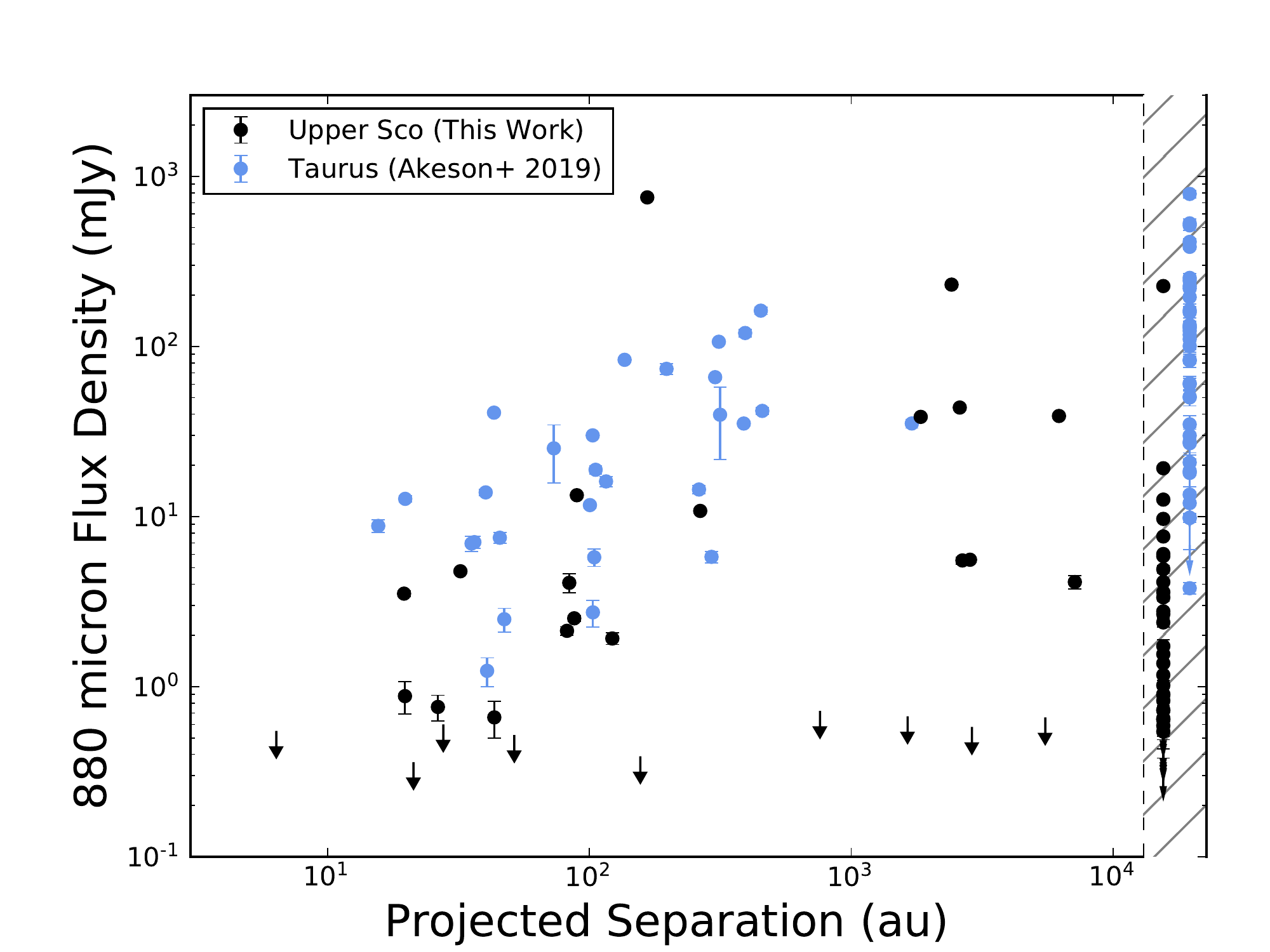}}
\caption{Total 880 $\mu$m continuum flux density and project companion separations of 
Upper Sco systems with primordial disks. Flux densities have been scaled to a common 
distance of 145 pc. Single stars are shown in the 
hatched region to the right of the figure. Taurus systems from \citet{Akeson2019} 
are shown in blue. Unlike in Taurus, where disks are significantly fainter 
in systems with companions, the brightness distributions of disks in systems 
with and without companions are indistinguishable in Upper Sco.
}
\label{fig:LmmSep}
\end{figure}

\begin{figure}[!h]
\centering
\subfloat{\includegraphics[width=0.50\linewidth]{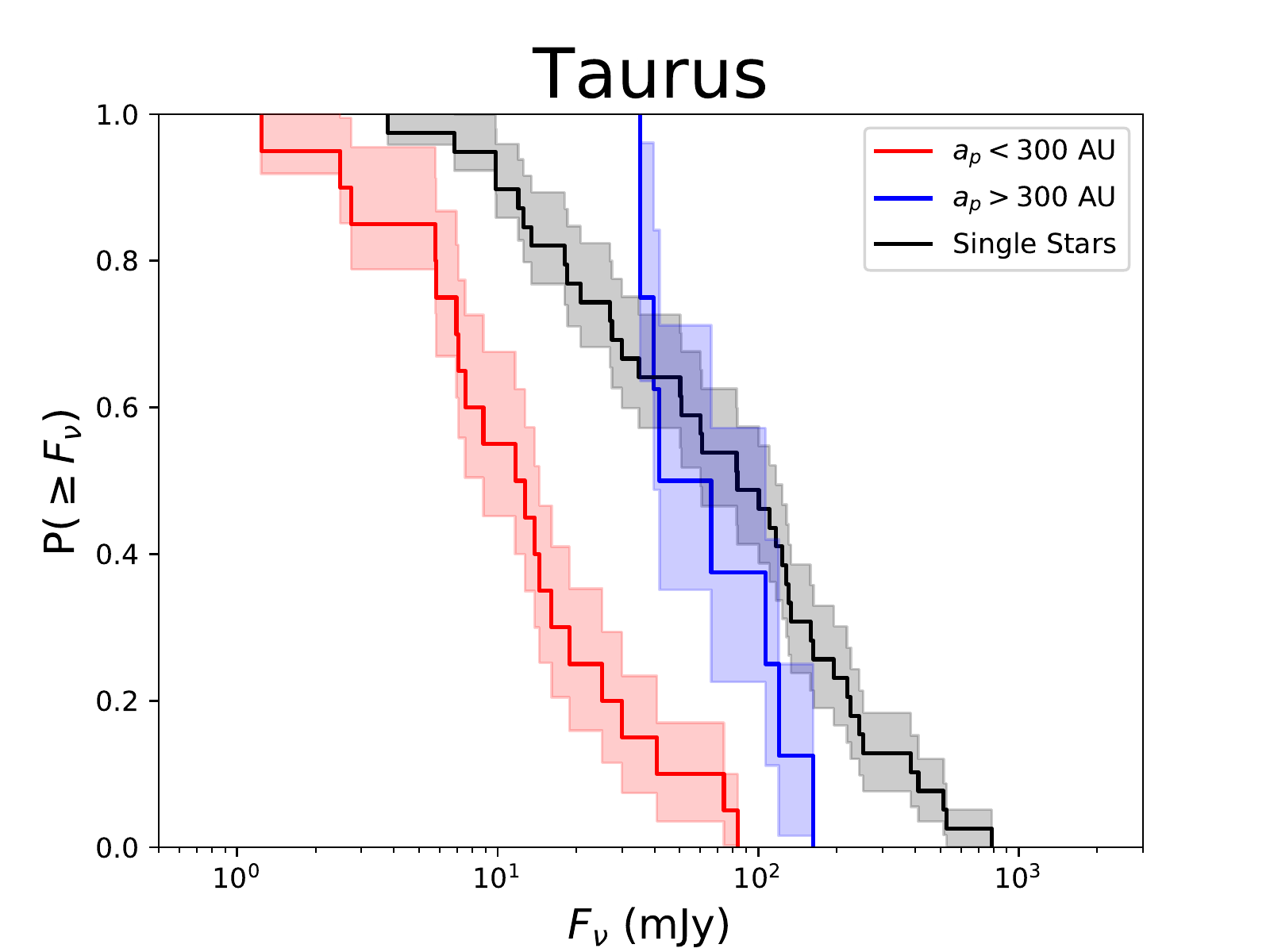}}
\subfloat{\includegraphics[width=0.50\linewidth]{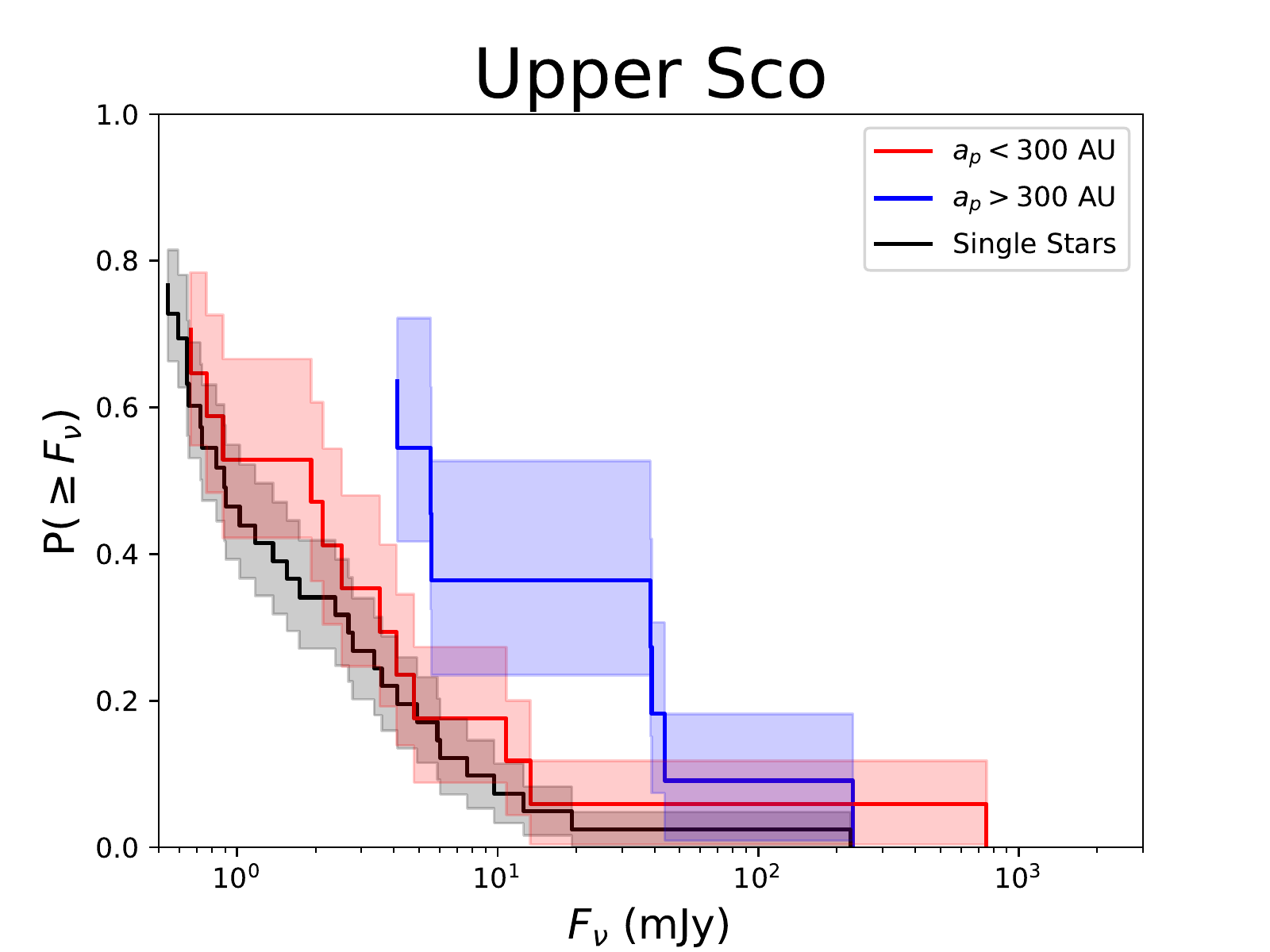}}
\caption{Cumulative distributions of 880 $\mu$m continuum flux density 
for the Taurus (left) and Upper Sco (right) systems shown in Figure 
\ref{fig:LmmSep}, calculated using the Kaplan-Meier product-limit estimator. 
Flux densities have been scaled to a common distance of 145 pc. 
In the case of Upper Sco, the distribution is only shown to the flux density 
of the faintest detection. Below this, the assumptions of the Kaplan-Meier 
product-limit estimator are violated, as all sources are upper limits.
In Taurus, single stars are significantly brighter than systems with companions 
within a projected separation of 300 au. In Upper 
Sco, however, the brightnesses are similar.
}
\label{fig:Surv_Curves}
\end{figure}

Figure \ref{fig:LmmSep} shows the 0.88 mm continuum flux densities of the binary and single systems in the Taurus and Upper Sco samples defined above.
Flux densities have been scaled to a common distance of 145 pc. 
Binaries are divided into systems with separation $<300$ au and $>300$ au, with flux densities representing the total emission of both components, following \citet{Harris2012}. 
We note that the Taurus and Upper Sco samples contain only eight and eleven systems, respectively, with separation $>300$ au. 
The flux density distinction between single stars and binaries separated by $<300$ au observed in Taurus is not present in Upper Sco. The difference is clearly 
apparent in Figure \ref{fig:Surv_Curves}, which
shows the cumulative flux distributions of single stars, systems with a companion beyond $300$ au, and systems with a companion within $300$ au. 
These distributions were calculated using the Kaplan-Meier product-limit estimator to account for the sources without a millimeter 
detection. 

In the case of Taurus, the flux distributions of systems with companions beyond $300$ au and single-star systems are statistically indistinguishable, 
with $p$-values of $0.50$ and $0.79$ given by the log-rank and Peto \& Peto Generalized Wilcoxian two-sample tests, implemented 
in \emph{R}. The brightnesses of the systems with companions within $300$ au are clearly lower, however. 
The log-rank and Peto \& Peto Generalized Wilcoxian two-sample tests
give $p$-values of $7.36\times10^{-5}$ and $4.22\times10^{-5}$ that these systems are drawn from the same brightness distribution as single stars. 
These results are consistent with those originally found by \citet[][see also Akeson et al. 2019]{Harris2012}.
We note that when comparing disk millimeter brightnesses, the observed correlation between disk brightness and stellar mass \citep{Andrews2013,Carpenter2014,Ansdell2016,Barenfeld2016,Pascucci2016,Ansdell2017}, must be taken into account. We find that the 
distributions of Taurus single star masses and primary masses for binaries with a separation of $<300$ au are 
statistically consistent with $p$-values of 0.69 and 0.62 given by the two versions of the Anderson-Darling test, implemented in $R$.
Therefore, the comparison of the disk luminosity distributions in these two samples are not affected by stellar mass bias.

For Upper Sco, the measured flux densities for single stars, wide companions, and close companions are all shifted to lower fluxes relative to Taurus. 
As with Taurus, 
the single-star and $>300$ au separation companion flux distributions 
are indistinguishable, with $p$-values of 0.43 and 0.10 given by the two-sample tests, although the sample size of wide companions 
is small. In contrast to Taurus, however, the flux distribution of $<300$ au companion systems is consistent with that of single stars 
in Upper Sco, with $p$-values of 0.85 and 0.62. 
As is the case for Taurus, stellar mass does not influence this result; the stellar mass distributions of Upper Sco stars with and 
without companions within 300 au are consistent, with $p$-values of 0.75 and 0.73.
Thus, it appears that while young disks 
in Taurus are strongly influenced by the presence of stellar companions, by the 5-11 Myr age of Upper Sco disk evolution has proceeded in such a way as to 
erase these initial effects.

\begin{figure}[!h]
\centerline{\includegraphics[scale=1.0]{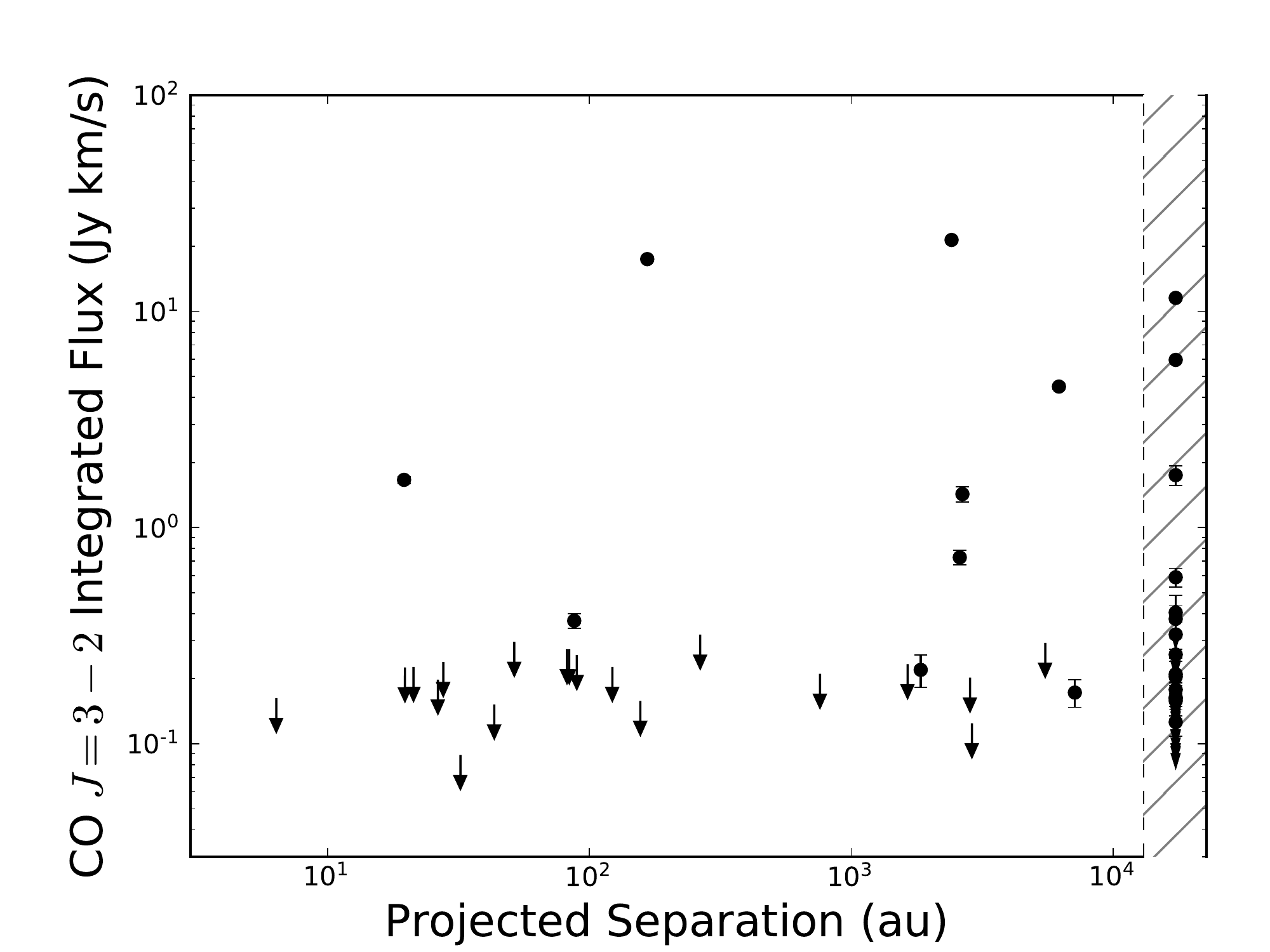}}
\caption{CO $J=3-2$ integrated line fluxes versus projected 
companion separations of Upper Sco systems with disks. 
Fluxes have been scaled to a common 
distance of 145 pc. Single stars are shown in the hatched region 
to the right of the figure. Although the distributions 
of fluxes for the single stars and systems with companions 
within 300 au are statistically indistinguishable, 11 out of 
37 single-star systems with fluxes below 0.5 Jy km s$^{-1}$ are detected, 
compared to none of the 14 such systems with companions separated by less
than 300 au. 
}
\label{fig:LCOSep}
\end{figure}

\begin{figure}[!h]
\centerline{\includegraphics[scale=1.0]{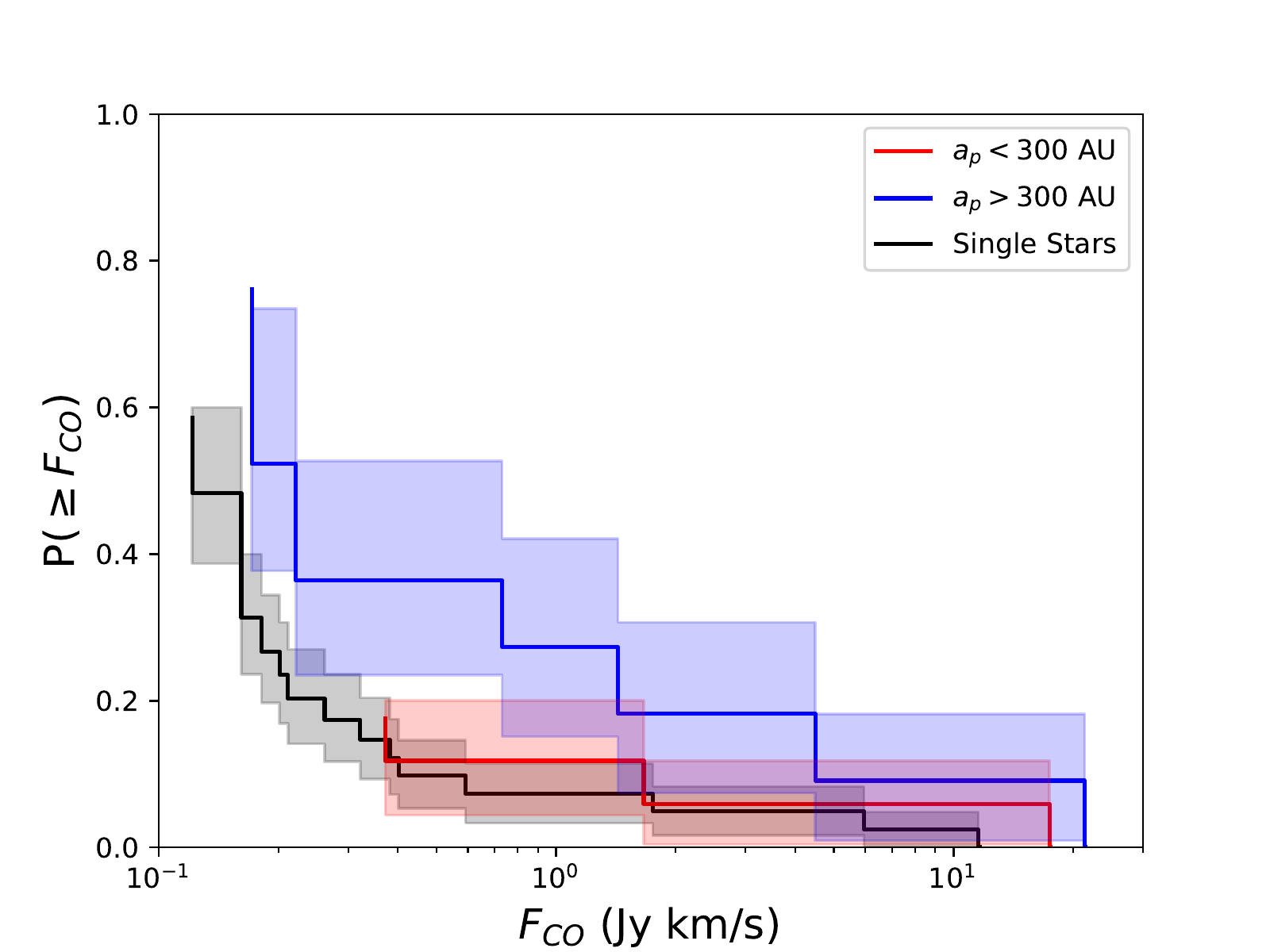}}
\caption{Cumulative distributions of the CO $J=3-2$ integrated line fluxes 
of Upper Sco systems with disks, calculated using the Kaplan-Meier product-limit 
estimator. Fluxes have been scaled to a common 
distance of 145 pc. The distribution is only shown to the flux 
of the faintest detection. Below this, the assumptions of the Kaplan-Meier 
product-limit estimator are violated, as all sources are upper limits.
The log-rank and Peto \& Peto Generalized Wilcoxian two-sample tests 
cannot distinguish between the flux distributions of single stars and systems with a 
companion within a projected separation of 300 au. 
}
\label{fig:CO_Surv_Curves}
\end{figure}

In Figure \ref{fig:LCOSep}, we compare the $^{12}$CO $J=3-2$ integrated line flux and projected separation for the Upper Sco primordial disks in binary systems. 
Figure \ref{fig:CO_Surv_Curves} shows the cumulative flux distributions, calculated using the Kaplan-Meier estimator as above. Both figures 
show similar CO flux distributions for single stars and systems with companions at CO fluxes greater than 0.5 Jy km s$^{-1}$, independent of companion separation. 
However, none of the 14 systems with a companion within $300$ au and CO flux below 0.5 Jy km s$^{-1}$ are detected in CO, while 11 of the 37 such single stars are detected. 
The Kaplan-Meier estimator is not reliable below 0.5 Jy km s$^{-1}$ due to the lack of detections in the former 14 systems. 
Therefore, the effects of binarity on gas and dust in disks may be different in Upper Sco. It is difficult to precisely 
quantify any such difference, however, as a 20\% reduction in the CO flux of these single systems would result in only three being detected, while a 30\% reduction 
would lead to none being detected. Thus, the lack of CO detections below 0.5 Jy km s$^{-1}$ in multiple systems may be due to only a small difference in flux.
Higher-sensitivity observations are necessary to definitively determine whether a difference exists in the CO integrated fluxes of disks with and 
without companions.

\subsection{Stellar Companions and Disk Evolution}

The observed correlation between the radial extent of millimeter-emitting grains and disk millimeter luminosity \citep{Barenfeld2017,Tripathi2017} 
suggests that 
the results of Section \ref{sec:Harris} can be 
explained by the evolution of dust disk sizes in single and multiple systems.
Disks in binary systems that are initially truncated by a stellar companion and survive to an age of 1-2 Myr will be smaller in size than their
counterparts in single-star systems. These truncated disks will thus be fainter, as is seen in Taurus \citep{Harris2012,Akeson2019}. A surface brightness comparison of disks in Taurus binary systems 
with those around single stars could measure the extent to which lower flux densities of binary system disks are due to 
this loss of the outer disk.

\citet{Barenfeld2017} measured the 
sizes of dust disks in Upper Sco, finding that these disks are smaller than younger systems by a factor of $\sim3$ on average. 
This suggests that the population of millimeter-sized grains in the outer disk is lost as disks evolve, providing a natural explanation for the similar luminosity distributions 
of disks in single and multiple systems in Upper Sco. Dust disks in multiple systems are truncated by their stellar companions, but their 
subsequent evolution is not as strongly affected by the presence of the companion after an age of 1-2 Myr, as shown in Section \ref{sec:Taurus_Comp}. Conversely, the outside-in evolution of single-star disks 
effectively allows them to ``catch-up'' to the smaller sizes of disks in multiple systems by an age of 5-11 Myr. The end result of dust disks tens of au in size with similar millimeter brightnesses 
is the same regardless of the presence or absence of a stellar companion. 

\citet{Gorti2015} have modeled disk evolution under the effects of viscous accretion, photoevaporation, dust radial migration, and dust growth 
and fragmentation, finding that the radial extent of millimeter-sized dust grains is expected to decline over time due to migration. The  
resulting millimeter dust disk sizes are similar to those measured for Upper Sco by \citet{Barenfeld2017}.  In this scenario, 
millimeter-sized grains from the outer disk replenish some of the dust lost from the inner disk due to viscous accretion, so that at the age of Upper Sco the inner disk is all that 
remains. However, disks in binary systems would lack this outer reservoir of millimeter-emitting grains due to tidal truncation, preventing the inner disk from being replenished.
This scenario would result in disks in binaries being fainter than disks in single systems, in contrast to what is observed in Section \ref{sec:Harris}. Thus, 
the shrinking of dust disks around single stars cannot simply be due to millimeter grains migrating inwards and remaining observable in the inner disk.

In addition to depletion through migration, the models of \citet{Gorti2015} predict that 
millimeter-sized grains will also be depleted in the outer disk through fragmentation. As photoevaporation lowers the density of gas 
in the outer disk, collisional velocities of millimeter dust grains will increase, leading to fragmentation into smaller grains 
that are not detectable at millimeter wavelengths. If this process occurs on a more rapid timescale than radial migration, it 
could provide a mechanism to remove outer disk millimeter grains without transporting them to the inner disk. Dust disks 
would therefore shrink in size without replenishment of the inner disk, resulting in disks having the 
same millimeter brightnesses in multiple and single systems.

On the other hand, local gas pressure maxima in disks are expected to trap concentrations of dust that would appear 
optically thick at millimeter wavelengths \citep[e.g.,][]{Whipple1972,Pinilla2012}. Optically thick dust substructure 
formed in this way has been suggested as an explanation of the observed correlation between dust disk size and millimeter luminosity 
seen in Taurus \citep{Tripathi2017} and Upper Sco \citep{Barenfeld2017}.
If millimeter grains in the inner disk are confined by dust traps to optically thick, unresolved 
substructures, dust could migrate into the inner disk without 
increasing its observed luminosity. Disks around single stars in Upper Sco would thus have higher dust masses than 
disks in multiple systems, but this extra material would be hidden by optical depth effects, causing single and 
multiple system disks to have the same millimeter brightnesses.

\section{Summary}
\label{sec:summary}
We have conducted a census of stellar companions around 112 stars with disks in the Upper Scorpius OB Association. Combining 
new observations with results from the literature, we find 30 sources brighter 
than $K=15$ and with separations of less than $2\arcsec$ from the target stars in 27 systems. These objects are likely to be companions based 
on the expected density of field stars. We compared the companion fraction of this sample to that of Upper Sco systems without disks \citep{Kraus2008} and 
investigated how the millimeter properties of these disks depend on companion separation.  The key conclusions of this 
paper are as follows.

\begin{enumerate}

\item ALMA images of the systems with disks and companions show that, for most such systems, the dust continuum emission 
is located around the primary or companion individually or is not detected toward either. For the systems with unresolved continuum 
emission encompassing both primary and companion, infrared SEDs show evidence for warm dust around one or both individual 
stars in the system.

\item Of the 50 primordial and debris/evolved transitional disk-hosting stars with spectral types G0-M3 in our sample, only seven have stellar 
companions brighter than $K=15$ with separations less than $2\arcsec$. 
Thirty-five systems in a comparison sample of 75 Upper Sco stars without disks in this spectral type range 
have stellar companions meeting the same brightness and separation criteria. 
The companion fraction for stars with disks is significantly lower, with a $p$-value 
of $2\times10^{-4}$. Restricting this comparison to primordial disks, we find that six of 26 stars with disks have 
a companion, a marginally lower fraction than that for stars without disks, with a $p$-value of 0.04.

\item The fraction of Upper Sco disk systems with a companion within 40 au is consistent with that of Taurus 
disks. While external stellar companions disrupt the early phases in disk evolution, as manifested in the lower disk fraction 
for close multiple systems than for single stars in Taurus, subsequent evolution appears to be dominated by internal 
disk processes. 

\item The observed distribution of millimeter continuum luminosity in Upper Sco is the same for disks in single-star systems and 
systems with a companion within a projected separation of 300 au. In contrast, disks in younger Taurus systems with such companions are fainter 
than those in single systems \citep{Harris2012,Akeson2019}, 
likely due to the smaller sizes of disks truncated by a stellar companion. 
This suggests that dust disks evolve from the outside-in between the ages of Taurus and Upper Sco, such that disks around single 
stars match the sizes and millimeter brightnesses of disks in binary systems by the 5-11 Myr age of Upper Sco.

\end{enumerate}

\acknowledgments
We thank the referee for their useful comments, which improved this manuscript.
We are grateful to Garreth Ruane, Ji Wang, and Henry Ngo for help reducing the NIRC2 imaging data. We thank Mike Ireland 
for use of his nonredundant aperture masking analysis code \linebreak(\url{https://github.com/mikeireland/idlnrm}) 
and Lynne Hillenbrand for valuable discussion regarding 2MASS J16075796-2040087.
This material is based upon work supported by the 
National Science Foundation Graduate Research Fellowship under grant No. DGE‐1144469.  S.A.B. acknowledges support 
from the NSF grant No. AST-1140063.  J.M.C. acknowledges 
support from the National Aeronautics and Space Administration under grant No. 15XRP15\_20140 issued through 
the Exoplanets Research Program. 
Some of the data presented herein were obtained at the W. M. Keck Observatory, which is operated as a scientific 
partnership among the California Institute of Technology, the University of California, and the National Aeronautics 
and Space Administration. The Observatory was made possible by the generous financial support of the W. M. Keck 
Foundation. The authors wish to recognize and acknowledge the very significant cultural role and reverence that the summit of 
Maunakea has always had within the indigenous Hawaiian community.  We are most fortunate to have the opportunity 
to conduct observations from this mountain. This research has made use of the Keck Observatory Archive (KOA), 
which is operated by the W. M. Keck Observatory and the NASA Exoplanet Science Institute (NExScI), under contract 
with the National Aeronautics and Space Administration. We thank Luca Rizzi for his aid with preparations for 
the NIRC2 observations and with telescope operation.
We are grateful to the ALMA staff for their assistance in the data
reduction. The National Radio Astronomy Observatory is a facility of the
National Science Foundation operated under cooperative agreement by Associated
Universities, Inc. 
ALMA is a partnership of ESO (representing its
member states), NSF (USA) and NINS (Japan), together with NRC (Canada) and NSC
and ASIAA (Taiwan), in cooperation with the Republic of Chile. The Joint ALMA
Observatory is operated by ESO, AUI/NRAO, and NAOJ.  
This work has made use of data from the European Space Agency (ESA) mission
\emph{Gaia} (\url{https://www.cosmos.esa.int/gaia}), processed by the \emph{Gaia}
Data Processing and Analysis Consortium (DPAC,
\url{https://www.cosmos.esa.int/web/gaia/dpac/consortium}). Funding for the DPAC
has been provided by national institutions, in particular the institutions
participating in the \emph{Gaia} Multilateral Agreement.
This publication makes use
of data products from the Two Micron All Sky Survey, which is a joint project
of the University of Massachusetts and the Infrared Processing and Analysis
Center/California Institute of Technology, funded by the National Aeronautics
and Space Administration and the National Science Foundation. 
This publication 
makes use of data products from the \emph{Wide-field Infrared Survey Explorer}, which 
is a joint project of the University of California, Los Angeles, and the Jet 
Propulsion Laboratory/California Institute of Technology, funded by the National 
Aeronautics and Space Administration. This work is based in part on observations 
made with the \emph{Spitzer Space Telescope}, which is operated by the Jet Propulsion Laboratory, 
California Institute of Technology under a contract with NASA.

Facilities: \facility{Keck:II (NIRC2)}, \facility{ALMA}, \facility{Gaia}, \facility{CTIO:2MASS}


\begin{figure}[!h]
\centering

\subfloat[][]{
\includegraphics[width=\textwidth]{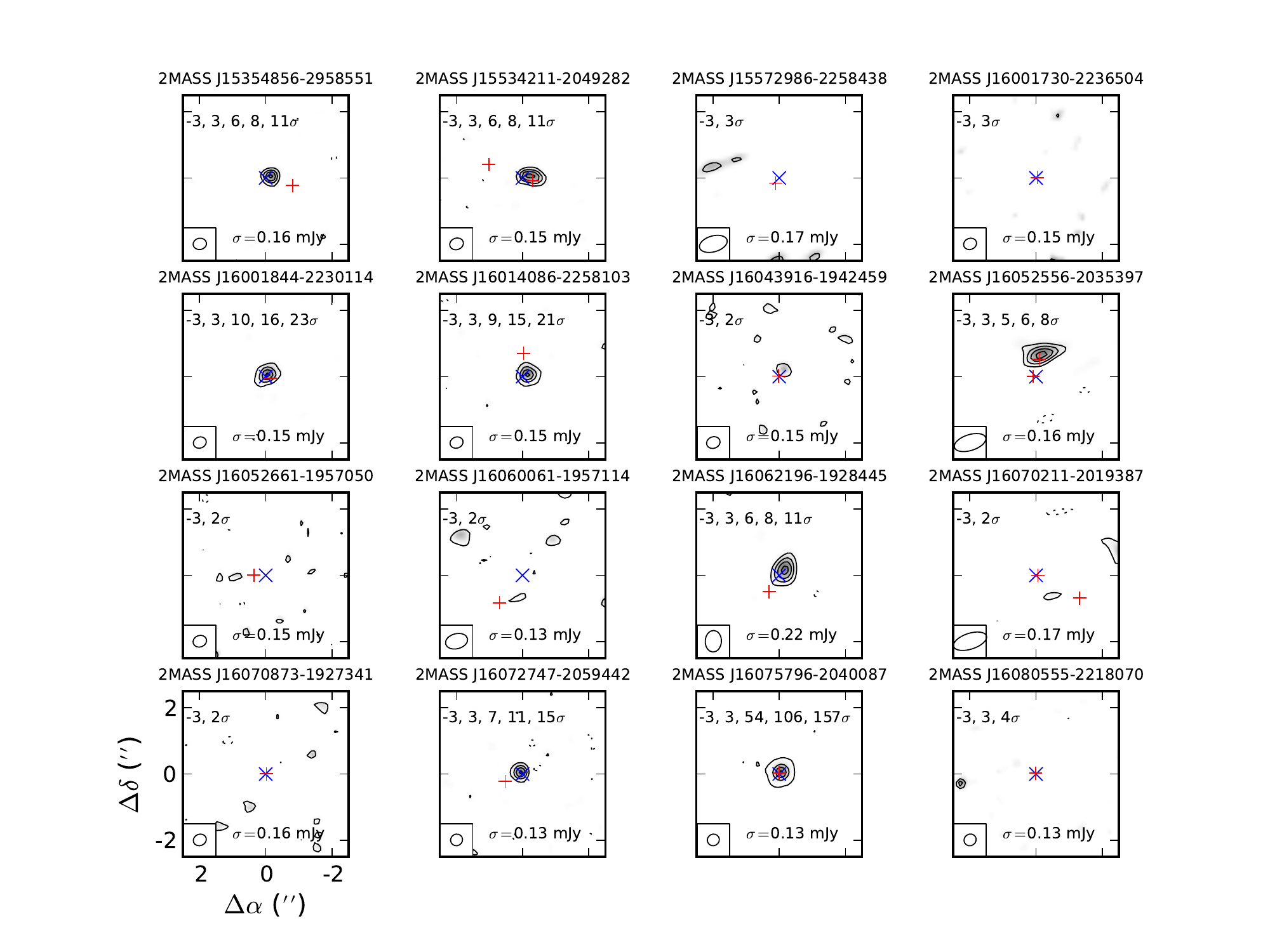}
}
\caption{ALMA 880 $\mu$m continuum images of the Upper Sco systems with disks and companions in this 
sample. These exclude 2MASS J16033471-1829303, which was not observed with ALMA. The relative 
positions of the primary (blue ``X'') and companion(s) (red ``+'') are overlaid. 
SEDs of the seven sources where the millimeter-wavelength emission cannot be 
conclusively assigned to the primary or secondary are shown in Figure \ref{fig:SEDs}.}
\label{fig:ALMA_Images}
\end{figure}

\begin{figure}
\ContinuedFloat
\centering
\subfloat[][]{
\includegraphics[width=\textwidth]{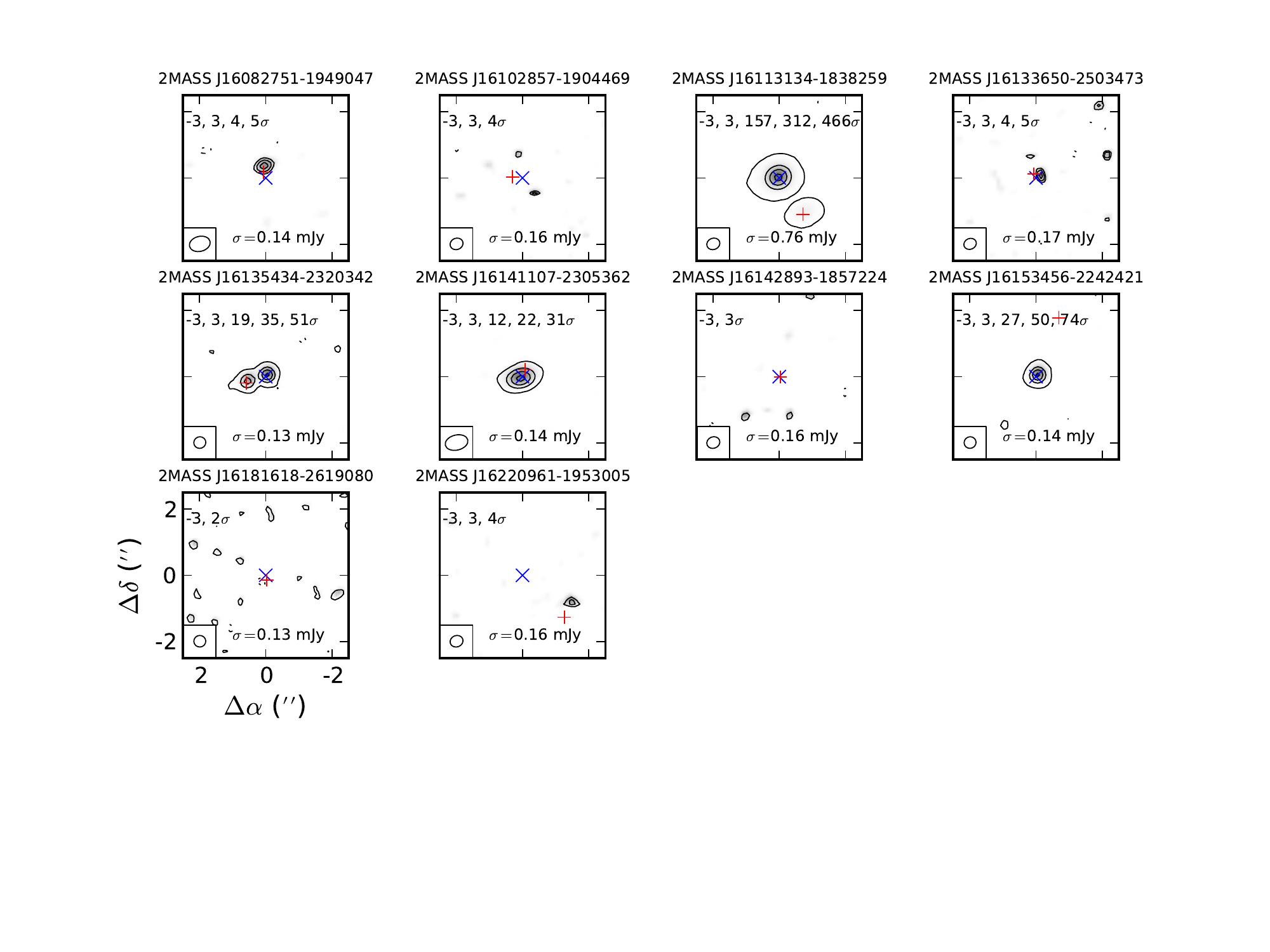}
}
\caption{Continued.}
\label{fig:ALMA_Images}

\end{figure}

\begin{figure}[!h]
\centering

\subfloat[][]{
\includegraphics[width=\textwidth]{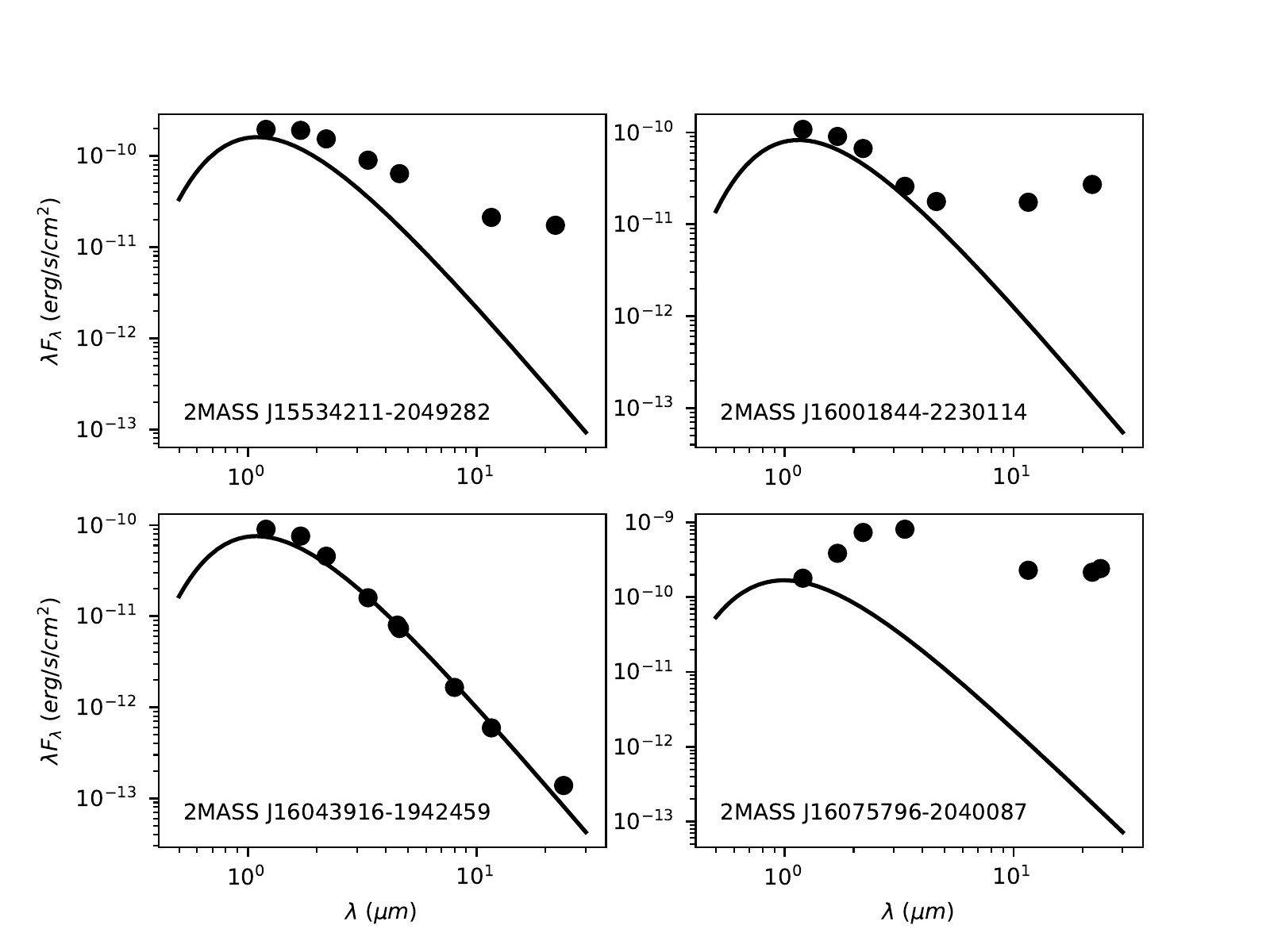}
}
\caption{Infrared SEDs of the systems in Figure \ref{fig:ALMA_Images} for which the millimeter-wavelength emission cannot be 
conclusively assigned to the primary or secondary. Stellar photospheric emission is estimated assuming 
blackbody emission with the stellar parameters calculated in \citet{Barenfeld2016}.
With the exception of 2MASS J16043916-1942459, all systems show excess at wavelengths 
$\leq8\mu$m, indicating that warm dust is present around the primary and/or companion(s) in these systems.
}
\label{fig:SEDs}
\end{figure}

\begin{figure}
\ContinuedFloat
\centering
\subfloat[][]{
\includegraphics[width=\textwidth]{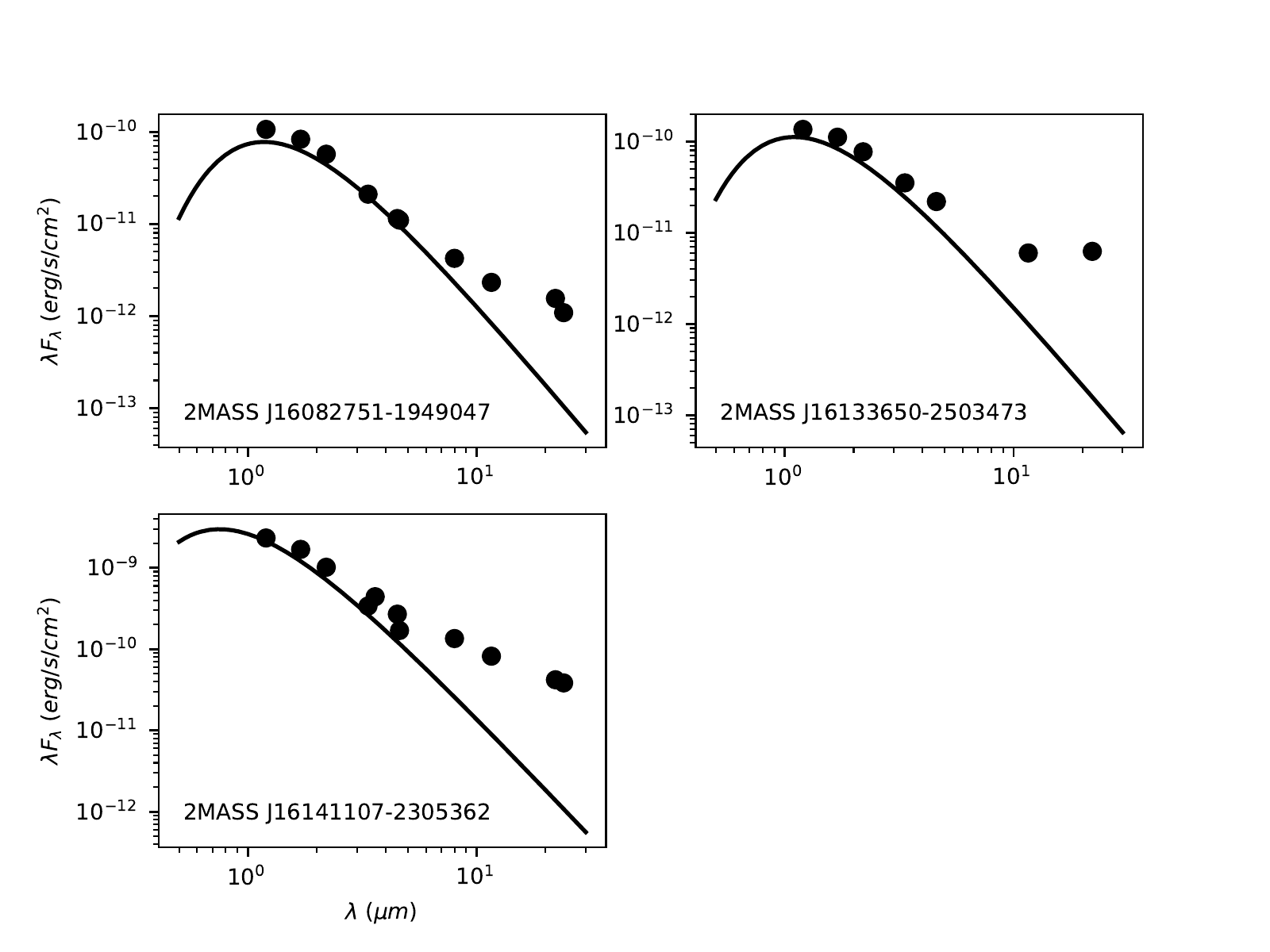}
}
\caption{Continued.}
\label{fig:SEDs}

\end{figure}

\newpage

\tabcolsep=0.15cm
\begin{deluxetable}{lccccccccc}
\tabletypesize{\tiny}
\tablecolumns{10} 
\tablewidth{0pt} 
\tablecaption{Upper Sco Disk Sample}
\tablehead{
  \colhead{Source}   &
  \colhead{Disk} &
  \colhead{Spectral}  &
  \colhead{$K$} &
  \colhead{Distance\tablenotemark{a}} &
  \colhead{$S_{0.88\mathrm{mm}}$\tablenotemark{b}} &
  \colhead{Observation} &
  \multicolumn{2}{c}{Integration Time (s)} &
  \colhead{Coronagraph?}\\

  \cline{8-9} 
  \colhead{} & 
  \colhead{Type} &
  \colhead{Type} &
  \colhead{(mag)} &
  \colhead{(pc)} & 
  \colhead{(mJy)} & 
  \colhead{Epoch\tablenotemark{c}} &
  \colhead{Imaging} &
  \colhead{Masking\tablenotemark{d}} &
  \colhead{}
}
\startdata
2MASS J15354856-2958551   &   primordial   		    &   M4     &   $9.46\pm0.03$	& $145\ (-11,+11)$ &  1.92  $\pm$  0.15	 & ...  	  &   ...   &  ...    &  ...  \\
2MASS J15514032-2146103   &   primordial   		    &   M4     &   $11.00\pm0.02$	& $142\ (-2,+2)$   &  0.76  $\pm$  0.16	 & 2013 May 30    &   60    &  120    &  no  \\
2MASS J15521088-2125372   &   primordial   		    &   M4     &   $12.08\pm0.03$	& $167\ (-7,+8)$   &  -0.10  $\pm$  0.15   & 2015 May 27    &   40    &  160    &  no  \\
2MASS J15530132-2114135   &   primordial   		    &   M4     &   $11.02\pm0.02$	& $146\ (-2,+3)$   &  5.78  $\pm$  0.14	 & 2013 May 30    &   60    &  120    &  no  \\
2MASS J15534211-2049282   &   primordial   		    &   M3.4   &   $9.62\pm0.03$	& $135\ (-3,+3)$   &  2.93  $\pm$  0.29	 & 2013 May 30    &   30    &  ...    &  no  \\
2MASS J15551704-2322165   &   debris/ev. trans.      	    &   M2.5   &   $9.33\pm0.02$	& $124\ (-2,+2)$   &  0.11  $\pm$  0.15	 & 2015 May 27    &   40    &  160    &  yes  \\
2MASS J15554883-2512240   &   debris/ev. trans.       	    &   G3     &   $8.29\pm0.02$	& $143\ (-1,+1)$   &  -0.14  $\pm$  0.15   & 2015 May 27    &   40    &  320    &  yes  \\
2MASS J15562477-2225552   &   primordial   		    &   M4     &   $10.79\pm0.02$	& $141\ (-2,+2)$ &    0.28  $\pm$  0.18	 & 2013 May 30    &   60    &  120    &  no  \\
\enddata
\tablenotetext{a}{Distances from \citet{BailerJones2018}. When no such distance was available, the mean and standard deviation of the rest of the sample, $145\pm11$ pc, were used.}
\tablenotetext{b}{0.88 mm continuum flux density measured by \citet{Barenfeld2016}. Ellipses indicate sources not observed with ALMA.}
\tablenotetext{c}{Ellipses indicate source with previous observation, summarized in Table \ref{tab:Lit_Comps}.}
\tablenotetext{d}{Ellipses in only this column indicate sources for which masking observations were not obtained due to the presence of a visual companion seen during observations.}
\tablenotetext{e}{Poor tip-tilt correction as discussed in Section \ref{sec:samp}}
\tablenotetext{f}{Target not visible in images as discussed in Section \ref{sec:samp}}
\label{tab:Disk_Sample}
\end{deluxetable}

\newpage
\begin{deluxetable}{lccccc}
\tabletypesize{\tiny}
\tablecolumns{6} 
\tablewidth{0pt} 
\tablecaption{Results from Previous Surveys}
\tablehead{
  \colhead{Primary}   &
  \colhead{Separation\tablenotemark{a}}  &
  \colhead{$\Delta K$\tablenotemark{b}} &
  \colhead{$K_{\mathrm{comp}}$\tablenotemark{c}} &
  \colhead{Position Angle\tablenotemark{d}} &
  \colhead{Reference} \\

  \colhead{} & 
  \colhead{(mas)} &
  \colhead{(mag)} & 
  \colhead{(mag)} &
  \colhead{(deg)} & 
  \colhead{} 
}
\startdata
2MASS J15354856-2958551   &   $844\pm3$     &   $0.09\pm0.08$      &   $9.55\pm0.10$   &   $254.40\pm0.03$    &  \citet{Kohler2000}    \\
2MASS J15583692-2257153   &   ...           &   ...    		   &   ...     &   ...   	     &  \citet{Kraus2008}    \\
2MASS J16001330-2418106   &   ...           &   ...    		   &   ...     &   ...   	     &  \citet{Kraus2008}    \\
2MASS J16014086-2258103   &   $706\pm1$     &   $0.84\pm0.03$      &   $10.68\pm0.06$   &   $357.5$\tablenotemark{e}   	     &  \citet{Bouy2006}    \\
2MASS J16020039-2221237   &   ...           &   ...    		   &   ...     &   ...   	     &  \citet{Kraus2008}    \\
2MASS J16025123-2401574   &   $7198\pm13$   &   $2.91\pm0.02$      &   $11.84\pm0.06$   &   $352.22\pm0.04$   &  \citet{Kraus2008}     \\
2MASS J16035767-2031055   &   ...           &   ...    		   &   ...     &   ...   	     &  \citet{Kraus2008}    \\
2MASS J16042165-2130284   &   ...           &   ...    		   &   ...     &   ...   	     &  \citet{Kraus2008}    \\
\enddata
\tablenotetext{a}{Ellipses indicate single stars. Contrast limits quoted by the previous studies, 
shown in Figure \ref{fig:limits} and listed in Table \ref{tab:limits}, are comparable to the current survey.}
\tablenotetext{b}{Difference in $K$ magnitude between primary and companion.}
\tablenotetext{c}{$K$ magnitude of companion.}
\tablenotetext{d}{Position angle is defined east of north.}
\tablenotetext{e}{Uncertainties not provided by authors.}
\label{tab:Lit_Comps}
\end{deluxetable}

\newpage
\begin{deluxetable}{lcccccc}
\tabletypesize{\small}
\tablecolumns{7} 
\tablewidth{0pt} 
\tablecaption{Newly Detected Sources}
\tablehead{
  \colhead{Primary}   &
  \colhead{Detection} & 
  \colhead{Separation}  &
  \colhead{$\Delta K$\tablenotemark{a}} &
  \colhead{$K_{\mathrm{comp}}$\tablenotemark{a}} &
  \colhead{Position Angle\tablenotemark{a}} &
  \colhead{Candidate} \\

  \colhead{Star} & 
  \colhead{Technique} &
  \colhead{(mas)} &
  \colhead{(mag)} & 
  \colhead{(mag)} &
  \colhead{(deg)} & 
  \colhead{Companion?\tablenotemark{b}} 
}
\startdata
2MASS J15530132-2114135   &   Imaging   &   $1690.2\pm11.4$  &   $7.13\pm0.10 $   &   $18.15\pm0.11$   &    $249.50\pm0.39$     &   no   \\
2MASS J15534211-2049282   &   Imaging   &   $321.8\pm0.1$    &   $1.47\pm0.01 $   &   $11.55\pm0.06$   &    $254.59\pm0.02$     &   yes   \\
2MASS J15534211-2049282   &   Imaging   &   $1097.1\pm0.1$   &   $1.47\pm0.01 $   &   $11.55\pm0.06$   &    $68.44\pm0.01$      &   yes   \\
2MASS J15554883-2512240   &   Imaging   &   $2306.2\pm0.8$   &   $5.87\pm0.01 $   &   $14.17\pm0.06$   &    $353.85\pm0.02$     &   no   \\
2MASS J15562477-2225552   &   Imaging   &   $2129.8\pm18.9$  &   $7.79\pm0.17 $   &   $18.58\pm0.18$   &    $59.77\pm0.51$      &   no   \\
2MASS J15570641-2206060   &   Imaging   &   $4381.7\pm39.1$  &   $6.90\pm0.08 $   &   $18.19\pm0.10$   &    $181.63\pm0.51$     &   no   \\
2MASS J15572986-2258438   &   Imaging   &   $194.4\pm1.5$    &   $0.26\pm0.06 $   &   $12.04\pm0.08$   &    $145.32\pm0.44$     &   yes   \\
2MASS J16001730-2236504   &   Masking   &   $43.1\pm1.4$     &   $1.70\pm0.06 $   &   $11.64\pm0.08$   &    $290.74\pm1.22$     &   yes   \\
\enddata
\tablenotetext{a}{$\Delta K$, $K_{\mathrm{comp}}$, and position angle are defined as in Table \ref{tab:Lit_Comps}.}
\tablenotetext{b}{Sources are considered to be candidate companions if they satisfy $K_{\mathrm{comp}}<15$ and separation $<2\arcsec$.}
\label{tab:companions}
\end{deluxetable}

\newpage
\begin{deluxetable}{lcccccccccc}
\tabletypesize{\tiny}
\tablecolumns{10} 
\tablewidth{0pt} 
\tablecaption{Contrast Limits for Systems without Detected Companions}
\tablehead{
  \colhead{}   &
  \colhead{} & 
  \multicolumn{9}{c}{$\Delta K$\tablenotemark{a}} \\

  \cline{3-11}
  \colhead{Primary} & 
  \colhead{Technique} & 
  \colhead{10-20} &
  \colhead{20-40} &
  \colhead{40-80} & 
  \colhead{80-160} &
  \colhead{160-240} & 
  \colhead{240-320}  &
  \colhead{320-500}  &
  \colhead{500-1000}  &
  \colhead{$>$1000}  
}
\startdata
2MASS J15514032-2146103   & Imaging      &   ...   &   ...   &   1.72   &   3.39   &   4.64   &   7.48   &   6.97   &   8.12   &   8.41    \\ 
2MASS J15514032-2146103   & Coronagraph  &   ...   &   ...   &   ...   &   ...   &   ...   &   ...   &   ...   &   ...   &   ...  \\ 
2MASS J15514032-2146103   & Masking      &   0.00   &   2.34   &   3.56   &   3.22   &   2.55   &   1.13   &   ...   &   ...   &   ...     \\

2MASS J15521088-2125372   & Imaging      &   ...   &   ...   &   ...   &   0.75   &   2.24   &   4.27   &   4.39   &   5.98   &   6.45    \\
2MASS J15521088-2125372   & Coronagraph  &   ...   &   ...   &   ...   &   ...   &   ...   &   ...   &   ...   &   ...   &   ...  \\
2MASS J15521088-2125372   & Masking      &   0.00   &   0.00   &   0.00   &   0.00   &   0.00   &   0.00   &   ...   &   ...   &   ...     \\

2MASS J15530132-2114135   & Imaging      &   ...   &   ...   &   0.81   &   3.27   &   4.45   &   7.33   &   6.88   &   7.78   &   8.26    \\
2MASS J15530132-2114135   & Coronagraph  &   ...   &   ...   &   ...   &   ...   &   ...   &   ...   &   ...   &   ...   &   ...  \\
\enddata
\tablenotetext{a}{Separation bins are reported in units of mas and $\Delta K$ in units of magnitude.}
\label{tab:limits}
\end{deluxetable}

\newpage
\begin{landscape}
\thispagestyle{empty}

\begin{deluxetable}{lcccccccccc}
\tabletypesize{\tiny}
\tablecolumns{11} 
\tablewidth{0pt} 
\tablecaption{Candidate Wide Companions from \emph{Gaia}}
\tablehead{
  \colhead{} & 
  \multicolumn{4}{c}{Primary} &
  \colhead{} &
  \multicolumn{5}{c}{Companion} \\

  \cline{2-5}
  \cline{7-11}
  \colhead{System}   &
  \colhead{$G$} & 
  \colhead{$\mu_{\alpha}$} &
  \colhead{$\mu_{\delta}$} &
  \colhead{Parallax} &
  \colhead{Separation} &
  \colhead{\emph{Gaia} Source Designation} & 
  \colhead{$G$} & 
  \colhead{$\mu_{\alpha}$} &
  \colhead{$\mu_{\delta}$} &
  \colhead{Parallax} \\

  \colhead{} & 
  \colhead{(mag)} &
  \colhead{(mas yr$^{-1}$)} &
  \colhead{(mas yr$^{-1}$)} & 
  \colhead{(mas)} &
  \colhead{($\arcsec$)} &
  \colhead{} &
  \colhead{(mag)} &
  \colhead{(mas yr$^{-1}$)} &
  \colhead{(mas yr$^{-1}$)} & 
  \colhead{(mas)} 
}
\startdata
2MASS J15551704-2322165   &    $8.306\pm0.001$   &   $-18.53\pm0.12$   &   $-28.11\pm0.07$   &   $8.24\pm0.10$   &   $14.8989\pm0.0807$   &  6237252529183461248  	&   $13.491\pm0.001$   &   $-19.66\pm0.20$   &   $-27.38\pm0.13$   &   $8.02\pm0.11$   \\
2MASS J15551704-2322165   &    $8.306\pm0.001$   &   $-18.53\pm0.12$   &   $-28.11\pm0.07$   &   $8.24\pm0.10$   &   $16.3749\pm0.1173$   &  6237252529183460992  	&   $14.150\pm0.002$   &   $-18.20\pm0.18$   &   $-26.98\pm0.12$   &   $8.02\pm0.09$   \\
2MASS J15551704-2322165   &    $8.306\pm0.001$   &   $-18.53\pm0.12$   &   $-28.11\pm0.07$   &   $8.24\pm0.10$   &   $17.9029\pm0.1522$   &  6237252529178690432  	&   $15.393\pm0.006$   &   $-18.51\pm0.29$   &   $-27.91\pm0.18$   &   $8.20\pm0.14$   \\
2MASS J15554883-2512240   &   $10.155\pm0.002$   &   $-15.85\pm0.10$   &   $-23.56\pm0.06$   &   $6.95\pm0.05$   &    $8.8551\pm0.1552$   &  6235742349962814592  	&   $13.918\pm0.001$   &   $-15.59\pm0.44$   &   $-22.23\pm0.26$   &   $8.99\pm0.22$   \\
2MASS J15554883-2512240   &   $10.155\pm0.002$   &   $-15.85\pm0.10$   &   $-23.56\pm0.06$   &   $6.95\pm0.05$   &   $14.5251\pm0.1366$   &  6235742349962814848  	&   $16.898\pm0.005$   &   $-15.64\pm0.31$   &   $-21.79\pm0.18$   &   $7.01\pm0.16$   \\
2MASS J16014086-2258103   &   $14.948\pm0.019$   &    $-8.80\pm0.26$   &   $-24.20\pm0.12$   &   $8.05\pm0.10$   &    $0.6950\pm0.0936$   &  6243135809748290688  	&   $15.357\pm0.001$   &   $-13.77\pm0.52$   &   $-26.85\pm0.20$   &   $9.19\pm0.43$   \\
2MASS J16025123-2401574   &   $11.866\pm0.003$   &   $-11.85\pm0.12$   &   $-24.03\pm0.05$   &   $6.95\pm0.07$   &    $7.2151\pm0.0837$   &  6236273895118890112  	&   $16.364\pm0.001$   &   $-12.70\pm0.32$   &   $-23.87\pm0.19$   &   $6.34\pm0.20$   \\
\enddata
\tablenotetext{a}{2MASS J16090002-1908368 and 2MASS J16090075-1908526 are a $18\farcs9511$ pair with matching parallaxes and proper motions.}
\label{tab:Gaia_Companions}
\end{deluxetable}
\end{landscape}

\newpage
\begin{deluxetable}{lcccccc}
\tabletypesize{\tiny}
\tablecolumns{7} 
\tablewidth{0pt} 
\tablecaption{Upper Sco Systems without Disks}
\tablehead{
  \colhead{Primary}   &
  \colhead{Spectral Type} & 
  \colhead{$K_{\mathrm{prim}}$\tablenotemark{a}}  &
  \colhead{Separation\tablenotemark{b}} &
  \colhead{$\Delta K$\tablenotemark{c}} &
  \colhead{$K_{\mathrm{comp}}$\tablenotemark{c}} &
  \colhead{Position Angle\tablenotemark{c}} \\

  \colhead{} & 
  \colhead{} &
  \colhead{(mag)} &
  \colhead{(mas)} & 
  \colhead{(mag)} &
  \colhead{(mag)} & 
  \colhead{(deg)} 
}
\startdata
2MASS J15355780-2324046   &   K3   &   $9.43\pm0.02$   &   $54.68\pm0.16$   &   $2.97\pm0.01$   &   $12.40\pm0.05$   &   $173.76\pm0.19$   \\
2MASS J15500499-2311537   &   M2   &   $8.93\pm0.02$   &   $26.93\pm0.04$   &   $0.76\pm0.04$   &   $9.69\pm0.07$   &   $222.07\pm0.11$   \\
2MASS J15505641-2534189   &   G0   &   $7.91\pm0.02$   &   $128\pm1$   	    &   $0.03\pm0.01$   &   $7.94\pm0.05$   &   $72.70\pm0.06$   \\
2MASS J15510660-2402190   &   M2   &   $9.73\pm0.02$   &   ...   &   ...   &   ...   &   ...   \\
2MASS J15545986-2347181   &   G3   &   $7.03\pm0.02$   &   $766\pm3$   	    &   $1.99\pm0.01$   &   $9.02\pm0.05$   &   $232.0\pm0.1$   \\
2MASS J15562941-2348197   &   M1.5 &   $8.75\pm0.02$   &   $92\pm6$   	    &   $0.62\pm0.05$   &   $9.37\pm0.07$   &   $169.8\pm5.0$   \\
2MASS J15565545-2258403   &   M0   &   $9.43\pm0.02$   &   ...   &   ...   &   ...   &   ...   \\
2MASS J15570234-1950419   &   K7   &   $8.37\pm0.02$   &   $558\pm1$   	    &   $0.54\pm0.01$   &   $8.91\pm0.05$   &   $292.1\pm0.3$   \\
\enddata
\tablenotetext{a}{Primary $K$ magnitude.}
\tablenotetext{b}{Ellipses indicate single stars.}
\tablenotetext{c}{$\Delta K$, $K_{\mathrm{comp}}$, and position angle are defined as in Table \ref{tab:Lit_Comps}.}
\label{tab:Kraus_Sample}
\end{deluxetable}

\end{document}